%% file: MAIN.tex
\documentclass{aa}  
\usepackage{natbib}
\bibpunct{(}{)}{;}{a}{}{,}
\usepackage{graphicx}
\usepackage[outercaption]{sidecap}
\usepackage{txfonts}
\usepackage[colorlinks=true, linkcolor=blue, citecolor=blue, urlcolor=magenta]{hyperref}
\usepackage{xcolor}

\newcommand{\mctwo}[1]{\multicolumn{2}{c}{#1}}

\begin{document} 
   \title{Probing the innermost regions of AGN jets and their magnetic fields with \textsl{RadioAstron}}
   \subtitle{IV. The quasar 3C\,345 at 18~cm: Magnetic field structure and brightness temperature\thanks{The reduced images (FITS format) are only available in electronic form at the CDS via anonymous ftp to cdsarc.u-strasbg.fr (130.79.128.5) or via \url{http://cdsweb.u-strasbg.fr/cgi-bin/qcat?J/A+A/}}}
   
   \author{F.~M.~Pötzl\inst{1}
          \and
          A.~P.~Lobanov\inst{1,2}
          \and
          E.~Ros\inst{1}
          \and
          J.~L.~G\'omez\inst{3}
          \and
          G.~Bruni\inst{4}
          \and
          U.~Bach\inst{1}
          \and
          A.~Fuentes\inst{3}
          \and
          L.~I.~Gurvits\inst{5,6,7}
          \and
          D.~L.~Jauncey\inst{7,8}
          \and
          Y.~Y.~Kovalev\inst{9,2,1}
          \and
          E.~V.~Kravchenko\inst{10,2,9}
          \and
          M.~M.~Lisakov\inst{1,9}
          \and
          T.~Savolainen\inst{11,12,1}
          \and
          K.~V.~Sokolovsky\inst{13,14}
          \and
          J.~A.~Zensus\inst{1}
          }

   \institute{Max-Planck-Institut für Radioastronomie,
             Auf dem Hügel 69, 53121 Bonn, Germany\\ \email{fpoetzl@mpifr-bonn.mpg.de} 
             \and
             Moscow Institute of Physics and Technology, 
             Institutsky per.~9, Dolgoprudny, Moscow region, 141700, Russia 
             \and
             Instituto de Astrof\'{\i}sica de Andaluc\'{i}a, CSIC, Apartado 3004, 18080, Granada, Spain 
             \and
             INAF - Istituto di Astrofisica e Planetologia Spaziali, via del Fosso del Cavaliere 100, 00133, Rome, Italy 
             \and
             JIVE - Joint Institute for VLBI ERIC, Oude Hoogeveensedijk 4, 7991 PD Dwingekoo, The Netherlands 
             \and
             Dept. of Astrodynamics and Space Missions, Delft University of Technology, Kluyverweg 1, 2629 HS Delft, The Netherlands 
             \and
             CSIRO Astronomy and Space Science, PO Box 76, Epping, NSW 1710, Australia 
             \and
             Research School of Astronomy and Astrophysics, Australian National University, Canberra, ACT 2611, Australia 
             \and
             Astro Space Center of Lebedev Physical Institute, Profsoyuznaya st.~84/32, Moscow, 117997, Russia 
             \and
             INAF Istituto di Radioastronomia, Via P. Gobetti 101, Bologna 40129, Italy 
             \and
             Aalto University Department of Electronics and Nanoengineering, PL 15500, FI-00076 Aalto, Finland 
             \and
             Aalto University Mets\"ahovi Radio Observatory, Mets\"ahovintie 114, FI-02540 Kylm\"al\"a, Finland 
             \and
             Center for Data Intensive and Time Domain Astronomy, Department of Physics and Astronomy, Michigan State University, 567 Wilson Rd, East Lansing, MI 48824, USA 
             \and
             Sternberg Astronomical Institute, Moscow State University, Universitetskii~pr.~13, 119992~Moscow, Russia\\ 
             }

\date{Received 22 September 2020 / Accepted 28 January 2021}

\abstract
   {Supermassive black holes in the centres of radio-loud active galactic nuclei (AGN) can produce collimated relativistic outflows (jets). Magnetic fields are thought to play a key role in the formation and collimation of these jets, but the details are much debated.}
   {We study the innermost jet morphology and magnetic field strength in the AGN 3C\,345 with an unprecedented resolution using images obtained within the framework of the key science programme on AGN polarisation of the Space VLBI mission \textsl{RadioAstron}.}
   {We observed the flat spectrum radio quasar 3C\,345 at 1.6~GHz on 2016 March 30 with \textsl{RadioAstron} and 18 ground-based radio telescopes in full polarisation mode.}
   {Our images, in both total intensity and linear polarisation, reveal a complex jet structure at 300~$\mu$as angular resolution, corresponding to a projected linear scale of about 2~pc or a few thousand gravitational radii. We identify the synchrotron self-absorbed core at the jet base and find the brightest feature in the jet 1.5~mas downstream of the core. Several polarised components appear in the Space VLBI images that cannot be seen from ground array-only images. Except for the core, the electric vector position angles follow the local jet direction, suggesting a magnetic field perpendicular to the jet. This indicates the presence of plane perpendicular shocks in these regions. Additionally, we infer a minimum brightness temperature at the largest $(u,v)$-distances of $1.1\times10^{12}$~K in the source frame, which is above the inverse Compton limit and an order of magnitude larger than the equipartition value. This indicates locally efficient injection or re-acceleration of particles in the jet to counter the inverse Compton cooling or the geometry of the jet creates significant changes in the Doppler factor, which has to be $>11$ to explain the high brightness temperatures.}
   {}


   \maketitle

\input{Sec1_Intro}

\input{Sec2_Obs}

\input{Sec3_Data_red}

\input{Sec4_Imaging}

\section{Results and discussion}

\input{Sec5_I_structure}

\input{Sec6_Tb}

\input{Sec7_P_structure}

\input{Sec8_Summary}

\begin{acknowledgements}
      We thank N.~R.~MacDonald and J.-Y.~Kim as well as the anonymous referee for valuable comments to the manuscript. The \textsl{RadioAstron} project is led by the Astro Space Center of the Lebedev Physical Institute of the Russian Academy of Sciences and the Lavochkin Scientific and Production Association under a contract with the State Space Corporation ROSCOSMOS, in collaboration with partner organizations in Russia and other countries. Partly based on observations performed with radio telescopes of IAA RAS (Federal State Budget Scientific Organization Institute of Applied Astronomy of Russian Academy of Sciences). The European VLBI Network is a joint facility of independent European, African, Asian, and North American radio astronomy institutes. Scientific results from data presented in this publication are derived from the following EVN project code(s): GG079A. Results of optical positioning measurements of the Spektr-R spacecraft by the global MASTER Robotic Net \citep{2010AdAst2010E..30L}, ISON collaboration, and Kourovka observatory were used for spacecraft orbit determination in addition to mission facilities. The National Radio Astronomy Observatory and the Green Bank Observatory are facilities of the National Science Foundation operated under cooperative agreement by Associated Universities, Inc. This research has made use of data from the MOJAVE database that is maintained by the MOJAVE team \citep{2018ApJS..234...12L}. Partly based on observations with the 100-m telescope of the MPIfR (Max-Planck-Institut f\"ur Radioastronomie) at Effelsberg. The data were correlated at the DiFX correlator \citep{2011PASP..123..275D, 2016Galax...4...55B} of the MPIfR at Bonn. A.P.L., Y.Y.K.\ and E.V.K.\ were supported by the Russian Science Foundation (project 20-62-46021).  L.I.G.\ acknowledges support by the CSIRO Distinguished Visitor Programme. T.S.\ was supported by the Academy of Finland projects 274477 and 315721. J.L.G\ acknowledges the support of the Spanish Ministerio de Econom\'{\i}a y Competitividad (grants AYA2016-80889-P, PID2019-108995GB-C21), the Consejer\'{\i}a de Econom\'{\i}a, Conocimiento, Empresas y Universidad of the Junta de Andaluc\'{\i}a (grant P18-FR-1769), the Consejo Superior de Investigaciones Cient\'{\i}ficas (grant 2019AEP112), and the State Agency for Research of the Spanish MCIU through the Center of Excellence Severo Ochoa award for the Instituto de Astrof\'{\i}sica de Andaluc\'{\i}a (SEV-2017-0709).
      This research has made use of NASA's Astrophysics Data System. This research has made use of adstex (\url{https://github.com/yymao/adstex}). This research has made use of the NASA/IPAC Extragalactic Database (NED), which is operated by the Jet Propulsion Laboratory, California Institute of Technology, under contract with the National Aeronautics and Space Administration.
\end{acknowledgements}

\bibliographystyle{aa} 
\bibliography{mybib.bib}

\end{document}

%% file: Sec1_Intro.tex
\section{Introduction}

A fraction of accreting supermassive black holes in the centres of active galactic nuclei (AGN) produce collimated relativistic outflows (jets) manifesting themselves through broadband continuum emission from the radio to gamma-ray regime. Blazars are a subclass of AGN, where the jet is closely aligned with the line of sight to the observer. Strong Doppler boosting in these sources makes them brighter and thus easier to detect compared to non-aligned AGN, and this makes them ideal laboratories for the studies of jets.

The physical processes of the formation of jets are still actively debated. Two promising jet launching mechanisms assume that either the jet is launched from the accretion disc \citep{1982MNRAS.199..883B} or from the rotating magnetosphere of the supermassive black hole itself \citep{1977MNRAS.179..433B}. The Event Horizon Telescope (EHT) observations of M87 \citep{2019ApJ...875L...5E} suggest that the jet is powered by magnetic fields anchored in the black hole, as postulated by the Blandford-Znajek mechanism. It might also be a combination of both launching mechanisms, as suggested from \textsl{RadioAstron} observations of 3C\,84 \citep{2018NatAs...2..472G}, which reveal a limb-brightened jet with a radius of 250~gravitational radii ($r_\mathrm{G}$) already at a distance of $350\,r_\mathrm{G}$ from the central engine. In either case, a dynamically important magnetic field is thought to play a crucial role in jet formation \citep{2001Sci...291...84M, 2014Natur.510..126Z}. The strength and morphology of the magnetic field within the innermost $10^4$ to $10^5$ $r_\mathrm{G}$ can therefore give crucial insight on how jets form and how they collimate and accelerate on parsec scales.

Often, the angular resolution of ground-based very long baselines interferometry (VLBI) arrays is insufficient to probe the innermost regions of distant AGN \citep{2020MNRAS.495.3576K}. With the inclusion of a space-borne radio telescope orbiting the Earth into the array, the maximum baselines can be extended to several Earth diameters ($D_\oplus$), effectively increasing the achieved angular resolution. The \textsl{RadioAstron} project \citep{2013ARep...57..153K} began its in-orbit operations in 2011, with the \textsl{Spektr-R} spacecraft launched on 2011 July 18, and remained operational until January 2019. It was equipped with a 10-m dish and had a high-eccentricity elliptical orbit with a major axis of $350000\,\mathrm{km}$, or $\sim\,27\,D_\oplus$, allowing for unprecedented $\mu$as resolution at observing frequencies of 0.32, 1.6, 4.8 and 22\,GHz. The space radio telescope (SRT) also provided, for the first time in Space VLBI, full polarisation capabilities, at 0.32, 1.6, and 22\,GHz.

The \textsl{RadioAstron} key science project (KSP) on AGN polarisation \citep[see Chapter 4.1 in][for the project description]{2020AdSpR..65..712B} aims to develop, commission, and exploit the unprecedented high angular resolution polarisation capabilities of \textsl{RadioAstron} to probe the innermost regions of AGN jets and their magnetic fields. With this it is possible to accurately determine potential Faraday rotation gradients \citep[e.g.][]{2011ApJ...733...11G, 2013MNRAS.436.3341Z} at the most compact angular scales, revealing changes of the magnetic field within the jet. Within the KSP, several of the brightest and also highly polarised AGN have been observed within the first four \textsl{RadioAstron} observing periods, AO-1, 2, 3 and 4, between 2013 and 2017. Results of these polarisation observations are reported for 0642+449 in \cite{2015A&A...583A.100L}, for BL~Lac in \cite{2016ApJ...817...96G} and 0716+714 in \cite{2020ApJ...893...68K}. \cite{2017A&A...604A.111B} analysed 3C\,273 in total intensity in paper II of the series.

Here we present the first full polarisation Space VLBI images of 3C\,345 (J1642+3948, 1641+399). The source is a compact, bright flat spectrum radio quasar (FSRQ) that has been observed with VLBI over several decades. The jet propagates at a viewing angle of $\sim 5^{\circ}$ \citep{2009A&A...507L..33P, 2012A&A...537A..70S} and exhibits apparent superluminal motions with speeds of $\sim 3$--20$\,c$ \citep[e.g.][]{1995ApJ...443...35Z, 2012A&A...537A..70S, 2019ApJ...874...43L}. The source underwent several flaring episodes in the optical, $\gamma$-rays and at radio wavelengths. \cite{2012A&A...537A..70S} were able to link the ejection of superluminal components to flaring events in 2009. Our observations are close to a local maximum in flux density of 3C\,345 (see the OVRO archive\footnote{\url{https://sites.astro.caltech.edu/ovroblazars/}}).

We assume a flat $\Lambda$CDM cosmology with $\Omega_m=0.3$ ,$ \Omega_\Lambda=0.7$, and $H_0=70\,\mathrm{km}\,\mathrm{s}^{-1}\mathrm{Mpc}^{-1}$ \citep{2014A&A...571A..16P}, so that 1\,milliarcsecond (mas) corresponds to 6.6\,pc projected distance for a redshift of $z=0.593$ ($D_\mathrm{A}=1.37\,\mathrm{Gpc}$) \citep{1996ApJS..104...37M} for 3C\,345.

%% file: Sec2_Obs.tex
\section{Observations}

\begin{figure}
    \centering
    \includegraphics[width=\hsize]{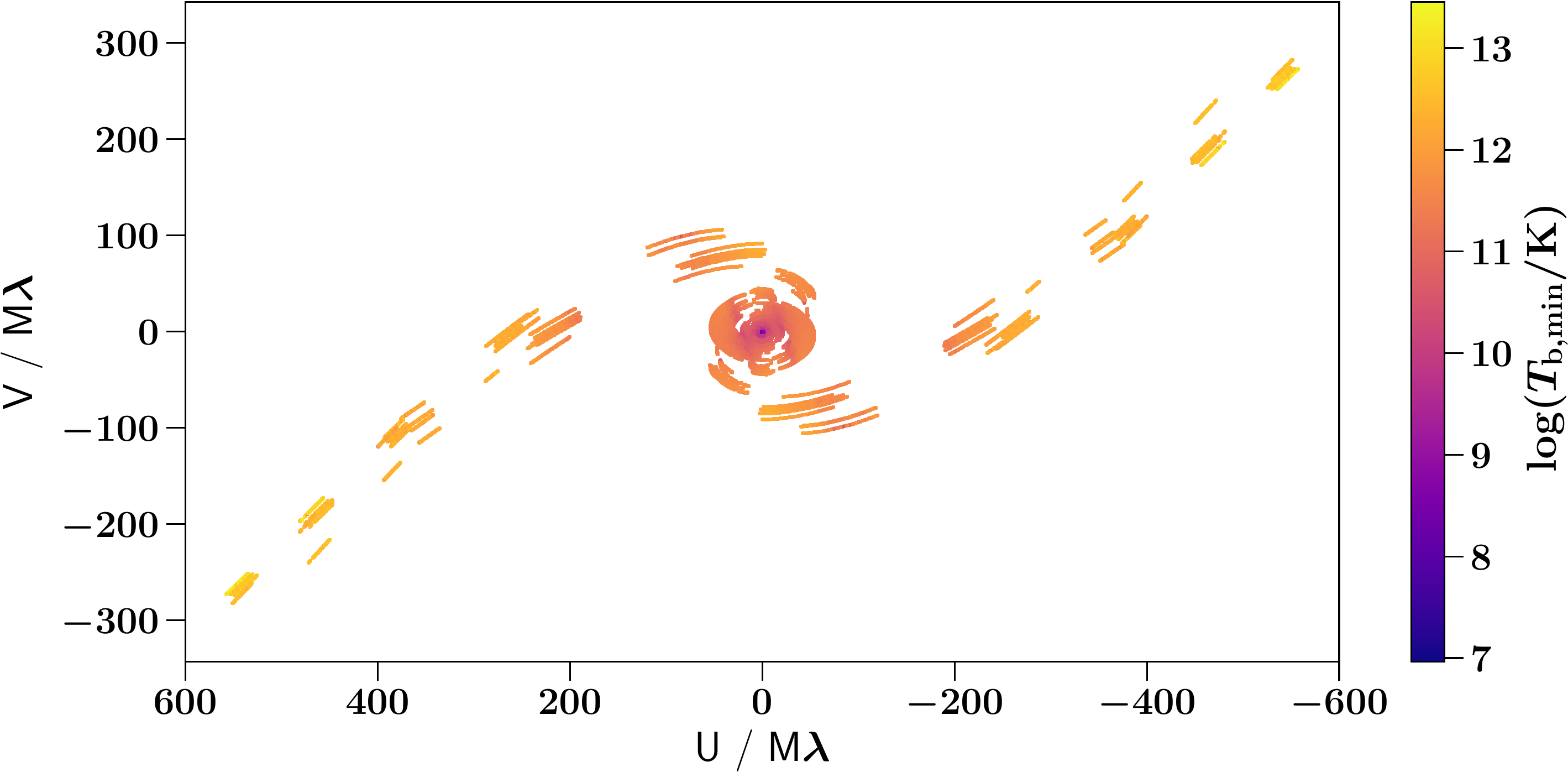}
    \caption{$(u,v)$-coverage of the observations of 3C\,345 described in this paper. The colour coding shows the minimum brightness temperature $T_\mathrm{b,min}$ estimated from the visibility amplitudes (see section \ref{sec:Tb} and \cite{2015A&A...574A..84L}).}
    \label{fig:uvcov_space}
\end{figure}

Observations of 3C\,345 at a central frequency of $1.6$\,GHz (18\,cm) were performed during AO-3 between 2016 March 30 21:00 UT and March 31 11:00 UT with an array of 18 antennas on the ground, complemented by the \textit{Spektr-R} spacecraft. The ground array consisted of the Very Long Baseline Array (VLBA) with Brewster (BR), Fort Davis (FD), Hancock (HN), Kitt Peak (KP), Los Alamos (LA), North Liberty (NL), Owens Valley (OV), Pie Town (PT) and Saint Croix (SC) (9 stations total), the Green Bank Telescope (GB), and eight European VLBI network (EVN) stations (Effelsberg (EF), Hartebeesthoek (HH), Jodrell bank (JB), Medicina (MC), Robledo (RO), Svetloe (SV), Torun (TR) and Zelenchukskaya (ZC)). Six more stations should have been observing, but had technical problems (Badary (BD), Mauna Kea (MK), Onsala (ON25), Sheshan (SH), Urumqi (UR) and Westerbork (WB1)). The resulting $(u,v)$-coverage from the remaining stations including the space baselines is shown in Fig.~\ref{fig:uvcov_space}.

The data were recorded in dual-polarisation mode (Right-hand circular (RCP) and Left-hand circular (LCP) polarisations), with four intermediate frequency bands (IFs) of 16\,MHz bandwidth each, yielding 64\,MHz total bandwidth for the ground stations, and two IFs for the space antenna, yielding 32\,MHz bandwidth. The data were correlated with the Space VLBI dedicated version of the DiFX software correlator, developed and run at the MPIfR in Bonn \citep{2016Galax...4...55B}.

%% file: Sec3_Data_red.tex
\section{Data reduction and calibration}

The data were calibrated using standard {\tt AIPS}\footnote{Astronomical Image Processing Software of the National Radio Astronomy Observatory, USA; \url{http://www.aips.nrao.edu/index.shtml}} \citep{2003ASSL..285..109G} procedures. The amplitudes were calibrated using the system temperatures ($T_\mathrm{sys}$) measured at the telescopes, where for SV and ZC we used median values due to sparse $T_\mathrm{sys}$ data. For RO, no $T_\mathrm{sys}$ measurements were available, so default values were used. Due to severe amplitude miscalibration, that could not be resolved later in the imaging process, stations JB and RO were dropped from further analysis, leaving us with 16 ground stations. For the SRT, the accuracy of the a priori amplitude calibration is considered at the level of 10-15\,\% \citep{2014CosRe..52..393K}. For the VLBA stations, an amplitude accuracy of $\sim 5\%$ can be expected \citep{2011A&A...532A..38S}. Typical amplitude errors of some other stations are given in \cite{2015A&A...583A.100L}, for example. For those stations with less accurate amplitude calibration, we used the well-calibrated antennas to gauge the overall calibration at the imaging stage. For the phase calibration (fringe-fitting), we performed a global antenna-based fringe-fitting with the task {\tt FRING}, applying an signal-to-noise ratio (S/N) threshold of 6. We tested whether the number and the S/N of solutions could be improved by first phasing up the ground array, and then using an exhaustive baseline search with baseline stacking while solving for the SRT. However, since it did not improve the fringe detection rate, we calibrated the whole array at once including the SRT. With this, ground-space fringes were found up to $9\,D_\oplus$. We did not observe decorrelation due to a possible time-dependent phase rate, that is to say the acceleration term was small throughout the experiment. The receiver bandpass was calibrated using standard AIPS routines, and for antennas with no good solutions for the bandpass, we flagged the outer 5 spectral channels on either side of IFs to minimise bandpass effects.

\subsection{Polarisation calibration}

The phase delay between RCP and LCP signals was calibrated using the task {\tt RLDLY} in {\tt AIPS}. After its application, a residual phase offset between RCP and LCP visibilities still remained. It was compensated by rotating the electric vector position angles (EVPAs) in the final polarisation map so that their directions, if convolved with a large beam, aligned with the corresponding vectors obtained in single dish data. For that purpose we used polarisation observations of the target source made with the Effelsberg Telescope at $1.6\,\mathrm{GHz}$ on the same day as our VLBI observations (U.~Bach, private communication). We believe that the single-dish observations reflect the EVPAs of the source on VLBI scales as it is very core dominated, with a core-to-extended flux ratio $>9$ based off VLA observations \citep{2004ApJ...608..698S}. The Effelsberg observations yield a total flux density of $S_\nu=6.77\pm0.14\,\mathrm{Jy}$, linearly polarised flux density $S_{\nu,P}=0.40\pm0.01\,\mathrm{Jy}$ ($S_{\nu,P}/S_\nu=5.92\pm0.23\,\%$) and $\chi=74\pm1\,^\circ$. Here and in the following, $P=\sqrt{Q^2+U^2}$ is the linearly polarised intensity calculated from Stokes $Q$ and $U$, $m=P/I$ is the fractional polarisation (total intensity denoted by Stokes $I$), and $\chi=0.5\times\arctan(U/Q)$ is the EVPA, measured from north to east.

The instrumental polarisation (the telescopes' $D$-terms) was calibrated using the {\tt AIPS} task {\tt LPCAL} and the total intensity (Stokes $I$) image of the source as input. The imaging procedure for the Stokes $I$ map is presented in Sect.~\ref{sec:imag}. The {\tt LPCAL} task assumes constant fractional polarisation for defined sub-components of the total intensity structure, where several sub-components were automatically generated with the task {\tt CCEDT} to reflect the complex source structure. The resulting $D$-terms were generally within $\sim10\,\%$ for all antennas, except for GB and SV. For GB a likely explanation for the poor $D$-term determination is the small parallactic angle coverage, as it only observed for a few scans. GB and SV were henceforth flagged out in the polarisation analysis because of the insufficient instrumental purity in our observations. Notably, we again confirm the polarisation capabilities of the SRT which demonstrated the instrumental polarisation of $\sim10\,\%$, in agreement with the previously reported values \citep{2015CosRe..53..199P, 2015A&A...583A.100L, 2016ApJ...817...96G}. For the SRT's LCP, we got $D$-term values of 6.6\,\% and 9.6\,\% for the two IFs, respectively. For RCP, we obtain $D$-term values as low as 3.0\,\% and 3.5\,\%, respectively. We used only the target source 3C\,345 for the $D$-term estimation. We found this sufficient since the main target was strong enough. The calibrator sources (3C\,286 and OJ\,287) were not well suited for $D$-term estimates since they were either observed only with a subset of telescopes with a sparse parallactic angle coverage and/or showed too much structure. Therefore we also do not provide error estimates of the $D$-terms.

\subsection{MOJAVE data}\label{sec:MOJ}

In addition to the \textsl{RadioAstron} data at 1.6\,GHz, we made use of archival MOJAVE\footnote{Monitoring Of Jets in Active galactic nuclei with VLBA Experiments} observations at 15\,GHz \citep{2018ApJS..234...12L}. These were conducted on 2016 March 5, less than a month apart from our \textsl{RadioAstron} observations, and will provide a reasonable comparison. This is justified by the observed variability in the radio light curves, and by the median velocity of jet components of $0.3\,\mathrm{mas}\,\mathrm{yr}^{-1}$. The 15\,GHz data are available at the MOJAVE webpage\footnote{\url{https://www.physics.purdue.edu/MOJAVE/sourcepages/1641+399.shtml}}.

\begin{figure}
    \centering
    \includegraphics[width=\hsize]{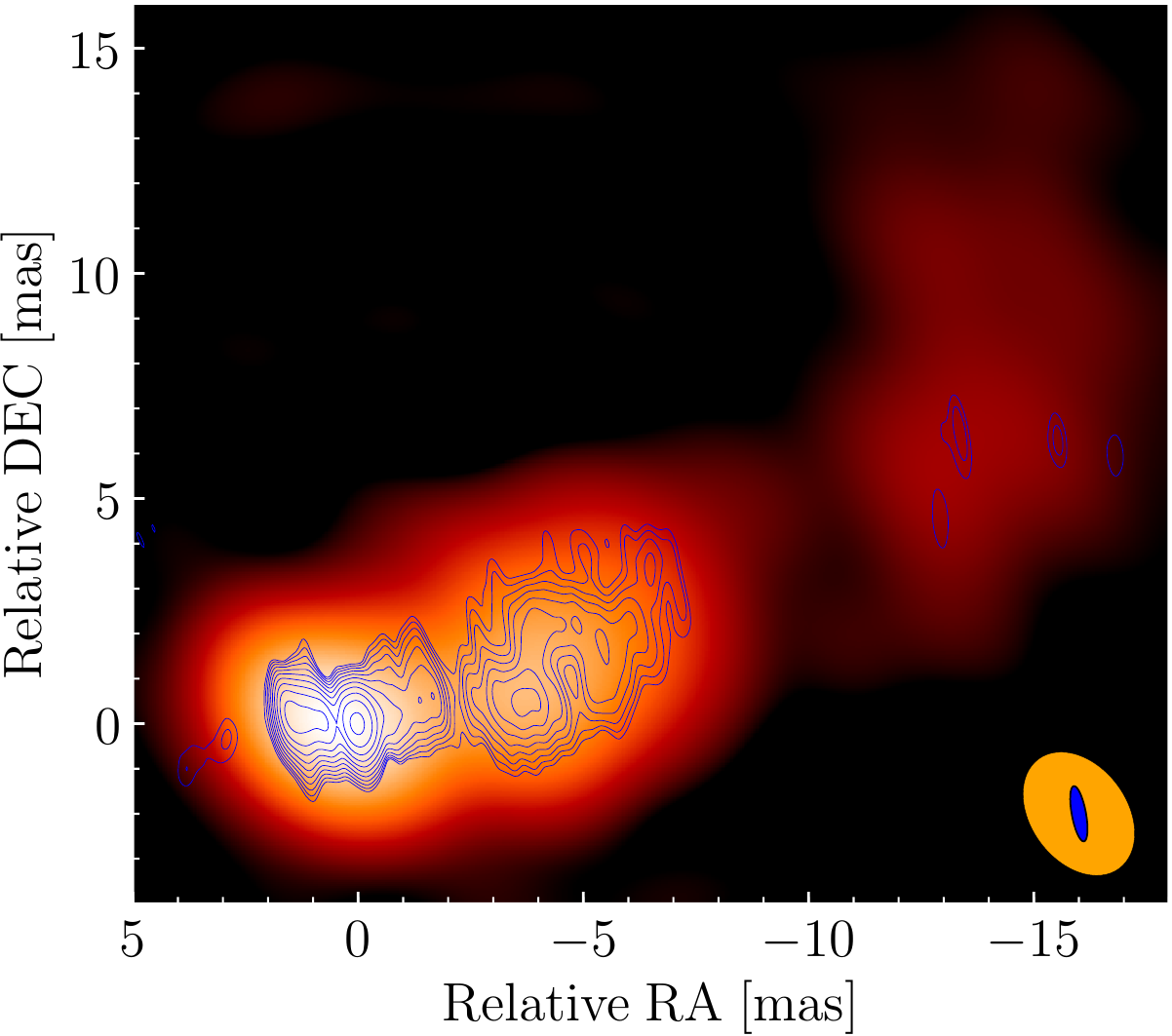}
    \caption{Total intensity image of 3C\,345 at 1.6\,GHz with the ground array data (orange scale) and all data including the space baselines (blue contours). The different beam sizes are displayed in the bottom right corner. We reach a resolution of $1.25\times0.32\,\mathrm{mas}$ with \textsl{RadioAstron}. Contour levels are in percents of peak emission of $0.39\,\mathrm{Jy/beam}$: 2.83, 4, 5.65, 8, 11.31, 16, 22.63, 32, 45.25, 64, 90.51. For the ground array image, the resolution is $3.0\times2.1\,\mathrm{mas}$, and the colour-scale shows the total intensity in log-scale between $0.5$ and $100\,\%$ of the peak of $1.68\,\mathrm{Jy}$.}
    \label{fig:I_maps_stack}
\end{figure}

\begin{figure}
    \centering
    \includegraphics[width=8cm]{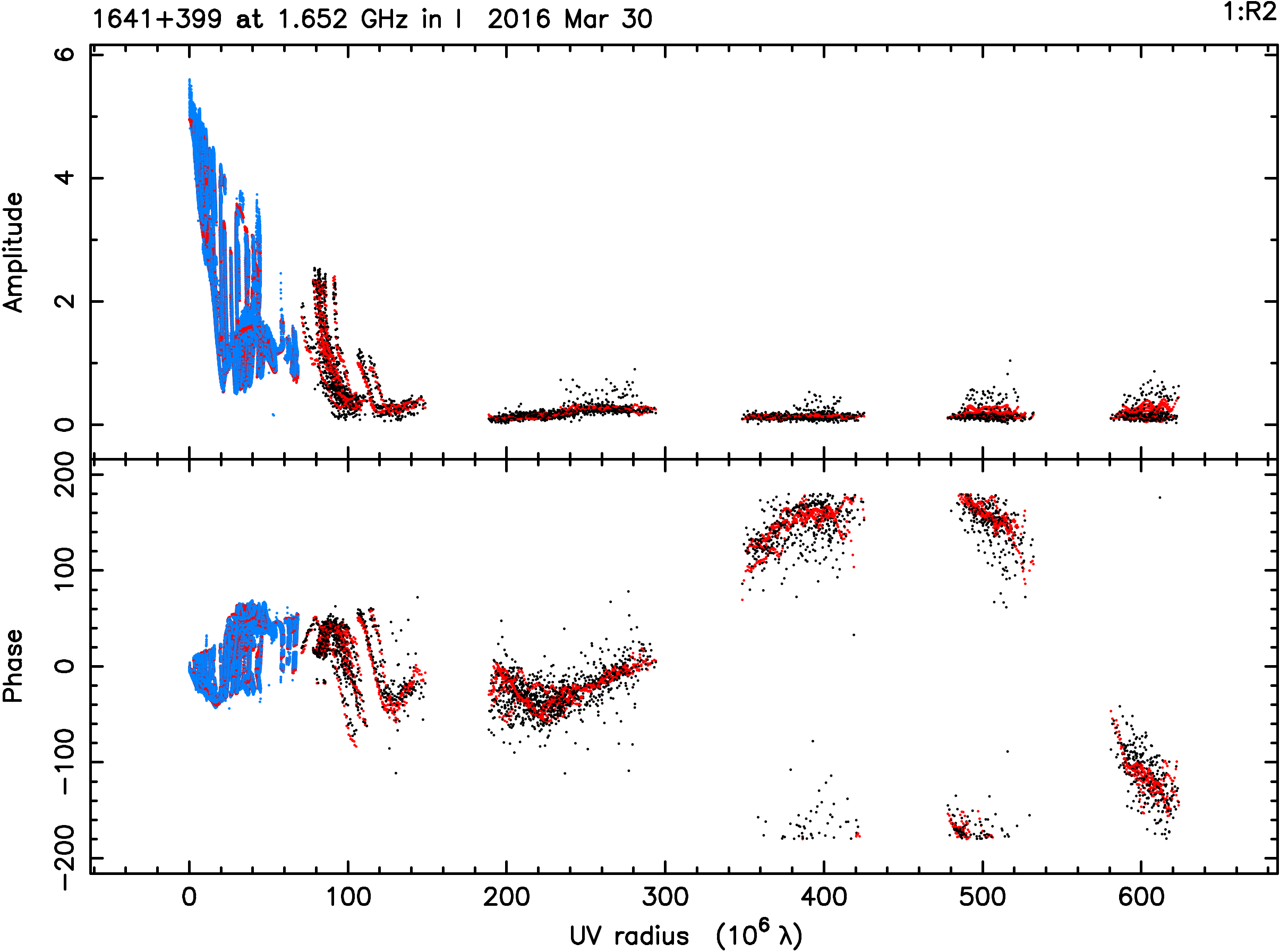}
    \caption{Visibility amplitudes (top) and phases (bottom) of the final calibrated data. Blue data points show ground only data, while black data points highlight the space baselines. The source \textsc{clean} model is shown in red.}
    \label{fig:radplot_RA}
\end{figure}
   
\begin{figure}
    \centering
    \includegraphics[width=8cm]{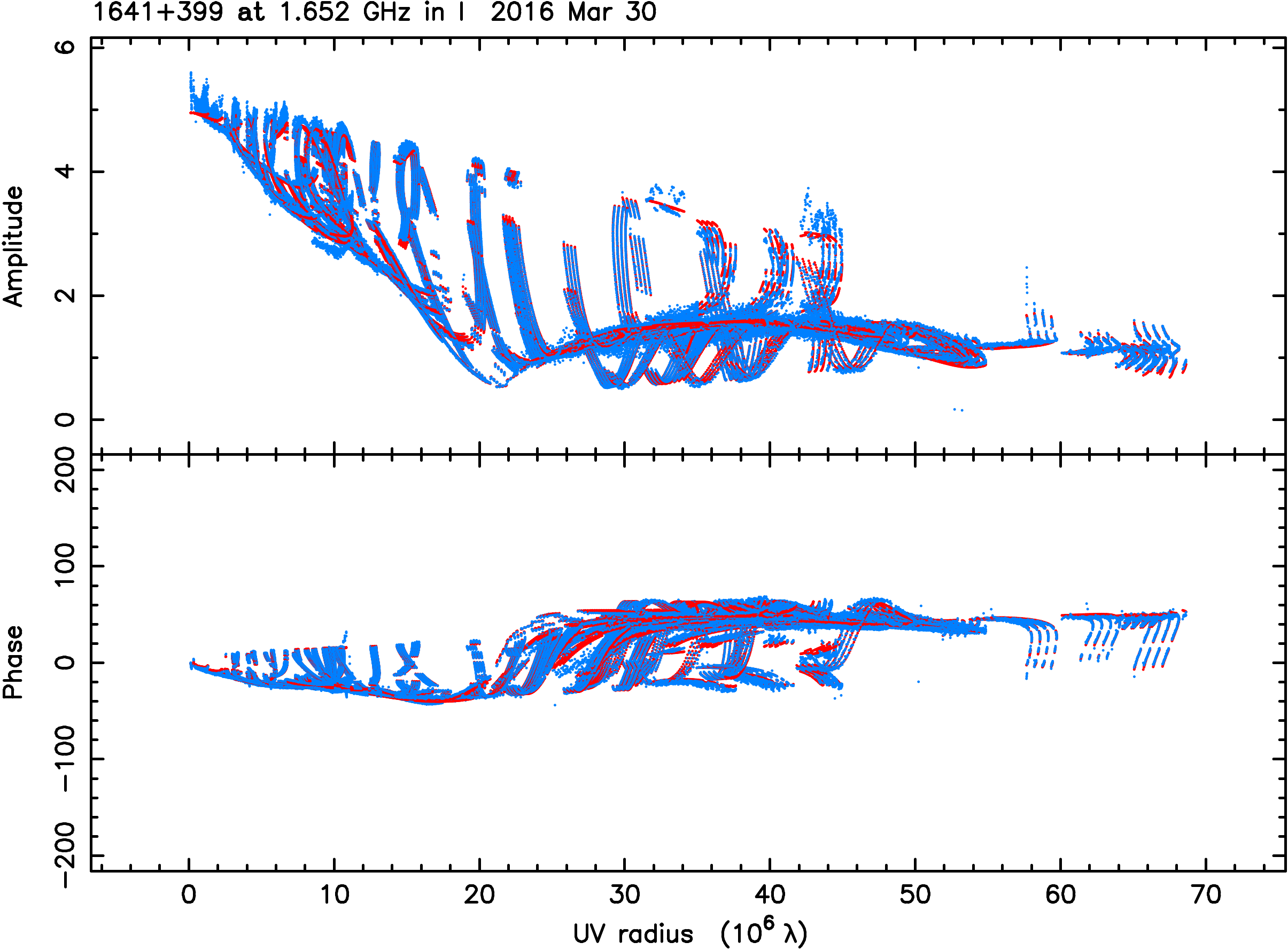}
    \caption{Visibility amplitudes (top) and phases (bottom) of data only from ground-based antennas. The source \textsc{clean} model is shown in red.}
    \label{fig:radplot_ground}
\end{figure}

%% file: Sec4_Imaging.tex
\section{Imaging} \label{sec:imag}

The data were imaged using the {\tt Difmap} software \citep{1997ASPC..125...77S}. Before imaging, we averaged the data into 90\,s intervals. We first created an image of the ground array only using standard \textsc{clean} and self-calibration procedures (see colour scale in Fig.~\ref{fig:I_maps_stack}). We then created a map using the baselines to the SRT based on the ground-only map, where we applied phase self-calibration down to a time interval of 3\,min and amplitude self-calibration only as overall gain factor to the SRT. The amplitude corrections we applied for the SRT were $8\,\%$ and $3\,\%$ for IF3 and IF4, respectively.

Fig.~\ref{fig:I_maps_stack} shows the images of 3C\,345 with the ground array only map in colour and with the full Space VLBI resolution in blue contours, using a uniform weighting scheme in both cases. The synthesised beam size is $1.25\times0.32\,\mathrm{mas}$ with \textsl{RadioAstron} and $3.0\times2.1\,\mathrm{mas}$ for the ground-array image. So the improvement in angular resolution of the obtained image due to the participation of \textsl{RadioAstron} over the ground-only image is about a factor of 7 along the jet direction. The image rms noise is $\sim1.7\,\mathrm{mJy/beam}$. The visibility amplitudes and phases as a function of projected $(u,v)$-distance are displayed in Fig.~\ref{fig:radplot_RA} and Fig.~\ref{fig:radplot_ground} for the whole array including the SRT and for the ground-array only data, respectively. An exemplary plot of the closure phases for the triangle R2-NL-ZC (R2 designating the SRT) is shown in Fig.~\ref{fig:cpplot} with the source model displayed in red.

The achieved resolution of $\sim300\,\mu\mathrm{as}$ (minor axis FWHM beam size) corresponds to a projected length of 2\,pc or between $\sim2600$ and $\sim10000\,r_\mathrm{G}$ for a black hole mass ranging between of $M_\mathrm{BH}\sim 2\times10^{9}\,\mathrm{M}_\odot$ and $M_\mathrm{BH}\sim 8\times10^{9}\,\mathrm{M}_\odot$ \citep{2001MNRAS.327.1111G, 2011ApJS..194...45S}.

%% file: Sec5_I_structure.tex
\subsection{Total intensity structure}\label{sec:I_str}

Our images reveal several components in the inner 10~mas of the jet, which could not be resolved with data from the ground array only. We find that the easternmost feature at the jet base is not the brightest component, which is a characteristic already seen in 3C\,345 with VSOP (VLBI Space Observatory Programme), the predecessor of \textsl{RadioAstron}  \citep{2000aprs.conf...21K, 2005ASPC..340...40K}. This feature likely corresponds to a partly synchrotron self-absorbed core, which we designate as the `core' in our subsequent analysis. We also observe a visibly curved jet structure in the few innermost mas of the jet, where the jet direction changes rapidly. We consider the weak easternmost feature of the jet visible at the edge of Fig.~\ref{fig:I_maps_stack} to be rather an imaging artefact than an indication of a counter-jet, as it is only about 3 times the noise level.

We fitted the visibilities with circular Gaussian components in the inner $8\,\mathrm{mas}$ of the jet using {\tt Difmap}. To find the minimum necessary number of components required to describe the structure, we used the criterion presented in \cite{2012A&A...537A..70S}. The fitted flux densities, positions, and sizes are listed in Table~\ref{tab:3C345_modelfit} and displayed in Fig.~\ref{fig:polmap}. The errors on those quantities were also calculated according to \cite{2012A&A...537A..70S}. The errors in polarised intensity, fractional polarisation and EVPA have been calculated from the map rms errors in Stokes $Q$ and $U$. In addition, we calculated the brightness temperature of each component, which we describe in detail in Sec.~\ref{sec:Tb}. A minimum possible size of a source structure which can still be resolved by an interferometer is dependent on the S/N, which is $\sim150$ for our 1.6\,GHz map. Combined with the restoring beam size of $1.25\times0.32\,\mathrm{mas}$, we estimate that features with an angular extent of
\begin{align}
    \theta_\mathrm{lim} = \dfrac{4}{\pi}\sqrt{\pi\log(2)b_\mathrm{maj}b_\mathrm{min}\log\left(\dfrac{\mathrm{S/N}}{\mathrm{S/N}-1}\right)}\sim100\,\mu\mathrm{as}
\end{align}
can be probed by our observations, according to \cite{2005astro.ph..3225L}. Our component sizes are all larger than this limit.

\subsection{Variability timescale}

\cite{2018Galax...6...49L} have investigated the variability properties of a large samples of AGN observed with \textsl{RadioAstron} in terms of their modulation index $\Bar{m}$ at 5\,GHz. 3C\,345 did not show any signs of intra-day variability (IDV), although the source is known to exhibit long-term (months to years) variability, as shown in observations at the Green Bank interferometer at 2 and 8\,GHz \citep{2006ApJS..165..439R}, with $\Bar{m}_{2\,\mathrm{GHz}}=0.012$, and the VLA MASIV program \citep{2008ApJ...689..108L}. Also at higher frequencies, up to 43\,GHz, 3C\,345 shows considerable variability \citep{2019A&A...626A..60A}. In addition, \cite{2019MNRAS.489.5365K} did not find signs for IDV at 15\,GHz, although \cite{2014MNRAS.438.3058R} found $\Bar{m}_{15\,\mathrm{GHz}}=0.129$ for long-term variability. The overall lack of IDV is not surprising considering the high Galactic latitude of 3C\,345, as IDV is likely caused by scintillation due to the Galactic interstellar medium \citep{2006ApJS..165..439R}. From our smallest component size (L3), we can estimate the shortest variability timescale according to \cite{2017ApJ...846...98J}:
\begin{align}
    \tau\sim\dfrac{25.3{\theta}D_\mathrm{L}}{\delta(1+z)}\,.
\end{align}
Here $\tau$ is the variability timescale in years, $\theta$ is the component FWHM in mas (as given in Table \ref{tab:3C345_modelfit}), $D_\mathrm{L}$ is the luminosity distance and $\delta$ is the Doppler factor. Considering Doppler factors between 10 and 25 (see Sect.~\ref{sec:Tb}), the shortest variability timescale is estimated to be between $1$ to $4$ months, where only the lower estimate is broadly consistent with $\tau=14.1$\,d at 2\,GHz, as found by \citep{2006ApJS..165..439R}. This indicates that the component size might still be slightly underestimated and not yet quite resolved. A size of the emitting region of about $100\,\mu\mathrm{as}$ is expected for sources with strong variability at higher frequencies, while not showing signs of scintillation due to the ISM \citep{2018MNRAS.474.4396K}.

\begin{table*}[thbp]
    \caption{Circular Gaussian model fit parameters and inferred brightness temperature from the \textsl{RadioAstron} and MOJAVE data.
    }
    \label{tab:3C345_modelfit}
    \centering
    \begin{tabular}{@{}c r@{$\pm$}l r@{$\pm$}l r@{$\pm$}l r@{$\pm$}l r@{$\pm$}l r@{$\pm$}l r@{$\pm$}l@{}}
        \hline\hline \noalign{\smallskip}
        (1)   & \mctwo{(2)}  & \mctwo{(3)}      & \mctwo{(4)}        & \mctwo{(5)}   & \mctwo{(6)}            & \mctwo{(7)}  & \mctwo{(8)} \\
        Comp. & \mctwo{Flux} & \mctwo{Distance} & \mctwo{P.A.}       & \mctwo{Size}  & \mctwo{$T_\mathrm{b}$} & \mctwo{$m$}  & \mctwo{$\chi$} \\
        & \mctwo{[mJy]} & \mctwo{[mas]}   & \mctwo{[$^\circ$]} & \mctwo{[mas]} & \mctwo{[K]}            & \mctwo{[\%]} & \mctwo{[$^\circ$]} \\ \noalign{\smallskip}
        \hline \noalign{\smallskip}
        \multicolumn{15}{c}{\textsl{RadioAstron} $1.6\,\mathrm{GHz}$}\\ \hline \noalign{\smallskip}
        Core & 382&45   & 1.48&0.07 &     79.0&2.7  & 0.37&0.04 & $(1.43$&$0.32)\times10^{12}$ & 1.6&0.7 &  43&6  \\
        L1   & 717&65   & 0.92&0.08 &     95.1&4.8  & 0.71&0.06 & $(7.16$&$1.33)\times10^{11}$ & 0.9&0.8 &  68&1  \\
        L2   & 471&37   & 0.19&0.04 &     44.8&12.4 & 0.20&0.01 & $(5.76$&$0.85)\times10^{12}$ & 6.4&0.6 &  56&1  \\
        L3   & 334&29   & 0.25&0.04 & $-112.4$&9.2  & 0.14&0.01 & $(8.96$&$1.38)\times10^{12}$ & 3.8&0.6 &  60&2  \\
        L4   & 373&30   & 1.09&0.07 &  $-76.6$&3.5  & 0.71&0.05 & $(3.72$&$0.60)\times10^{11}$ & 4.7&2.4 &  55&7  \\
        L5   & 182&24   & 1.76&0.12 &  $-46.0$&4.0  & 0.73&0.09 & $(1.71$&$0.49)\times10^{11}$ & 7.6&3.3 &  51&6  \\
        L6   & 1068&108 & 3.73&0.19 &  $-79.3$&2.9  & 1.85&0.18 & $(1.59$&$0.35)\times10^{11}$ & 6.7&1.4 &  94&3  \\
        L7   & 831&144  & 5.69&0.42 &  $-67.6$&4.2  & 2.36&0.40 & $(7.59$&$2.87)\times10^{10}$ & 4.4&3.4 & 153&11 \\ 

        \noalign{\smallskip} \hline \noalign{\smallskip}
        \multicolumn{15}{c}{MOJAVE $15\,\mathrm{GHz}$} \\ \hline \noalign{\smallskip}
        Core & 1445&38 & 0.083&0.01  &     91.9&6.5 & \mctwo{$<0.02$} & \mctwo{$>1.94\times10^{13}$} & 2.87&0.04 &   55.0&0.1 \\
        U1   & 2200&47 & 0.077&0.01  &  $-87.9$&5.8 & 0.08&0.01       & $(2.19$&$0.08)\times10^{10}$ & 3.64&0.03 &   55.0&0.2 \\
        U2   & 312&18  &   0.34&0.02 &  $-93.2$&4.0 & 0.14&0.01       & $(9.24$&$1.01)\times10^{10}$ &  4.3&0.1  &   50.7&0.4 \\
        U3   & 455&22  &  0.871&0.02 & $-111.9$&1.3 & 0.14&0.01       & $(1.33$&$0.11)\times10^{11}$ & 13.1&0.3  &   19.1&0.3 \\
        U4   &  97&11  &   1.39&0.08 &  $-95.4$&3.4 & 0.55&0.06       & $(1.92$&$0.44)\times10^{9}$  &  6.5&1.6  & $-9.9$&3.1 \\
        U5   & 353&20  &   1.84&0.02 & $-106.9$&0.7 & 0.18&0.01       & $(6.81$&$0.67)\times10^{10}$ &  5.9&0.4  &   12.4&0.8 \\
        U6   & 286&23  &   2.43&0.06 &  $-88.9$&1.5 & 0.78&0.06       &  $(2.75$&$0.47)\times10^{9}$ &  9.1&1.4  &   26.7&2.3  \\
        U7   & 188&22  &   5.37&0.14 &  $-89.9$&1.5 & 1.27&0.15       &  $(6.92$&$1.78)\times10^{8}$ &  9.6&3.3  &   66.1&4.9  \\
        U8   & 379&54  &   6.59&0.35 &  $-77.9$&3.0 & 2.50&0.35       &  $(3.58$&$1.12)\times10^{8}$ &  8.6&7.3  &   80.4&10.5 \\ 
        \hline \noalign{\smallskip}
        \multicolumn{15}{p{0.85\textwidth}}{\tiny \textsc{Note:}  Columns display the (1) component name (see Fig.~\ref{fig:polmap} and \ref{fig:polmap_MOJ}), (2) flux density, (3) radial distance from the total intensity peak, (4) component position angle, (5) component FWHM size, (6) brightness temperature, (7) fractional polarization and (8) Electric Vector Position Angle. Where the component size was smaller than the minimal resolvable source size in the map, we provide an upper limit. Errors are purely statistical errors and may be underestimated. See text for details.
        }
        \end{tabular}
\end{table*}

\begin{figure}
    \centering
    \includegraphics[width=\hsize,angle=0]{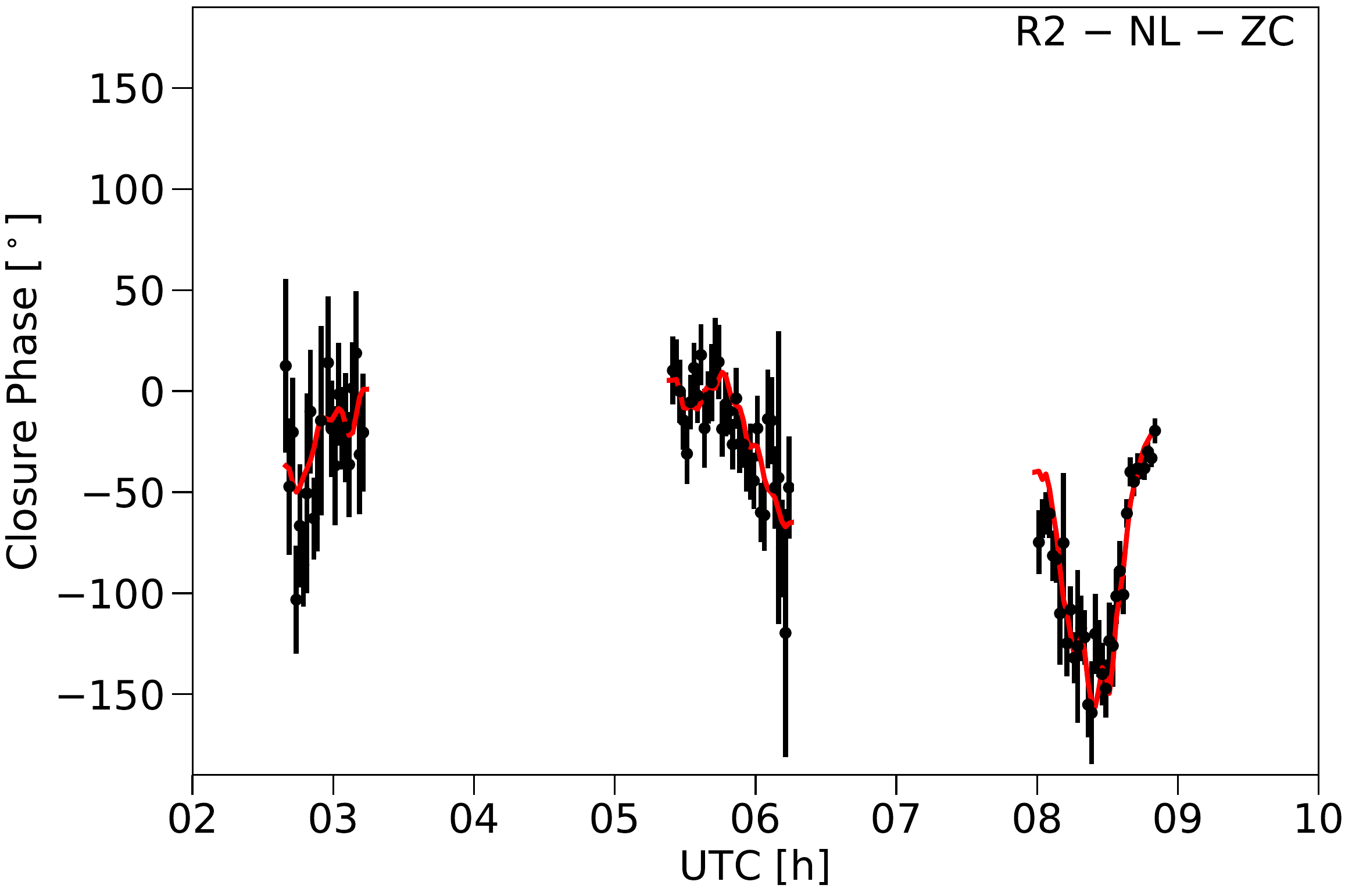}
    \caption{Closure phases of the triangle R2-NL-ZC, where R2 designates the SRT. The source clean model is shown in red. The \textsl{RadioAstron} perigee occurred during the last scans.
    }
    \label{fig:cpplot}
\end{figure}

%% file: Sec6_Tb.tex
\subsection{Brightness temperature}\label{sec:Tb}

\begin{SCfigure*}
    \caption{Minimum brightness temperature $T_\mathrm{b,min}$ and maximum brightness temperature $T_\mathrm{b,max}$ as a function of $(u,v)$-distance. The values were estimated from the visibilities following \cite{2015A&A...574A..84L}. The solid (dashed) lines show the average $T_\mathrm{b,min}$ ($T_\mathrm{b,max}$) in bins of $10\,\mathrm{M}\lambda$ for \textsl{RadioAstron} (black) and MOJAVE (red), while the points show the individual data values with the same colour scheme. The dashed horizontal lines show $T_\mathrm{b}$ calculated from Gaussian modelfits (see Table~\ref{tab:3C345_modelfit}). $T_\mathrm{b,min}$ and $T_\mathrm{b,max}$ provide a reasonable bracketing for the brightness temperature at least for the \textsl{RadioAstron} data.\vspace{1.1cm}}\hspace*{8pt}
    \includegraphics[width=14cm]{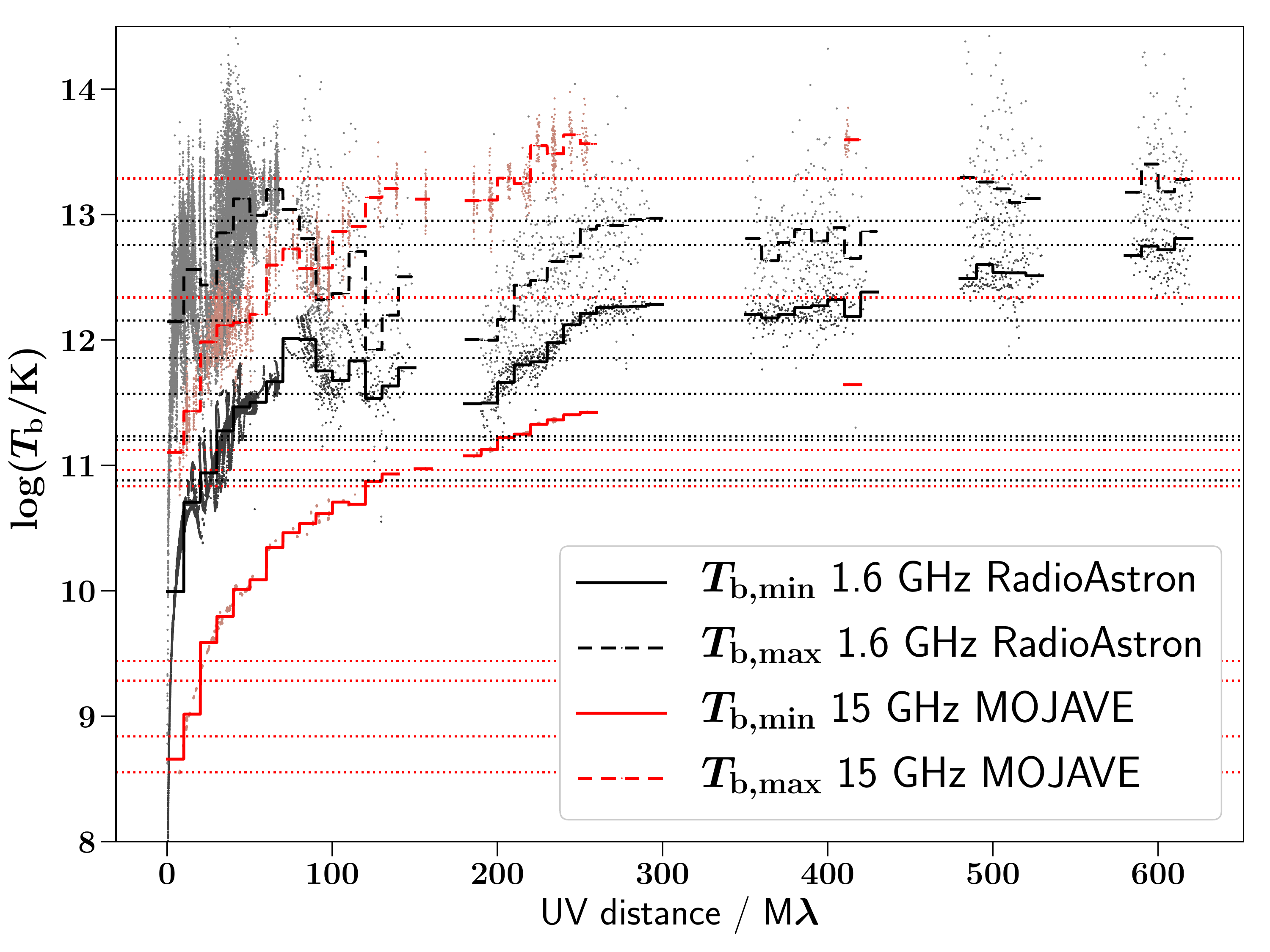}
    \label{fig:UV_Tb_minmax}
\end{SCfigure*}

The ability of an interferometer to measure the brightness temperature is in principle independent from the observing wavelength, and depends only on the projected interferometer baseline \citep{2005AJ....130.2473K}. Accordingly, \textsl{RadioAstron} uniquely probes the highest brightness temperatures \citep[e.g.][]{2016ApJ...817...96G,2016ApJ...820L...9K,2018MNRAS.475.4994K,2018MNRAS.474.3523P,2020AdSpR..65..705K,2020ApJ...893...68K}.

We calculate the brightness temperature in two ways. We first estimate the minimum brightness temperature $T_\mathrm{b,min}$ as well as the maximum brightness temperature $T_\mathrm{b,max}$ from the visibility data and the visibility errors according to \cite{2015A&A...574A..84L}. Calculating the minimum brightness temperature requires the assumption of a circular or axially symmetric brightness distribution. For the maximum brightness temperature, in addition, one has to assume that the structure at the probed scale is marginally resolved. Following that, we calculate the brightness temperature from the fitted flux densities and FWHM of the modelfit components explained in Sect.~\ref{sec:I_str}. The values of the estimated brightness temperature from both methods are presented in Fig.~\ref{fig:UV_Tb_minmax}, as well as in the $(u,v)$-coverage plot in Fig.~\ref{fig:uvcov_space} for the first method. We also did the same calculations for the archival MOJAVE observations described in Sect.~\ref{sec:MOJ}. We go into more detail about the comparison of the data sets in Sect.~\ref{sec:Tb_MOJ}.

The estimates of the brightness temperature from the visibilities are most accurate for baselines $>200\,\mathrm{M}\lambda$, where $T_\mathrm{b,min}$ and $T_\mathrm{b,max}$ provide a reliable constraint on $T_\mathrm{b}$ \citep{2015A&A...574A..84L}. As shown in Fig.~\ref{fig:UV_Tb_minmax}, we binned the data into  $10\mathrm{M}\lambda$ bins, and for the bin at the largest $(u,v)$-distances we get an average $T_\mathrm{b,min}=6.46\times 10^{12}\,\mathrm{K}$ and $T_\mathrm{b,max}=1.90\times 10^{13}\,\mathrm{K}$, which are the values that we use in our following analysis.

\begin{SCfigure*}
    \includegraphics[width=140mm]{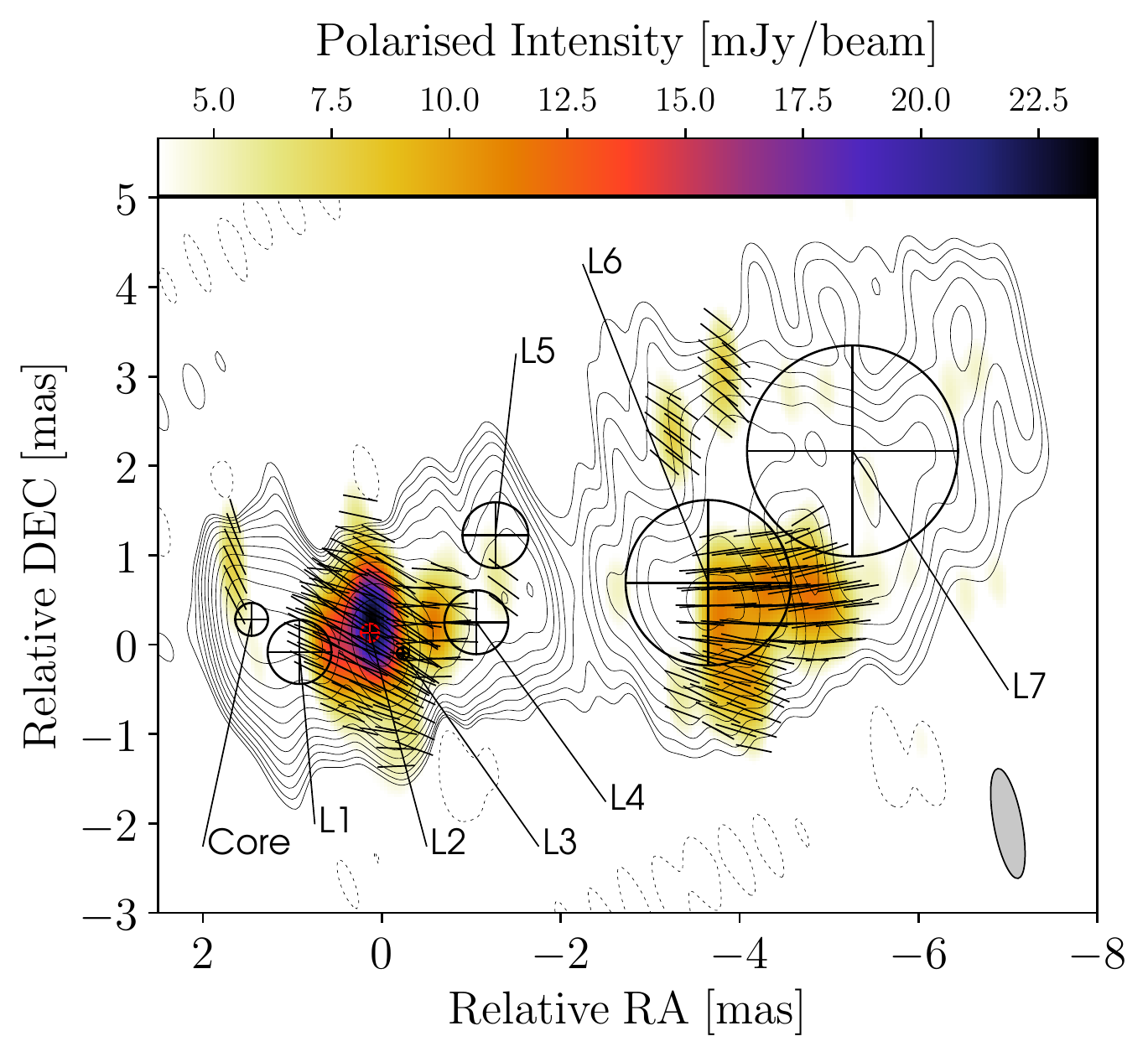}
    \caption{\textsl{RadioAstron} image of the total intensity and polarised emission of 3C\,345 at 1.6\,GHz. Map of the polarised intensity $P$ in colour-scale, overlaid with contours displaying the total intensity emission. The beam size is displayed on the bottom right with a resolution of $1.25\times0.32\,\mathrm{mas}$. The lines show the EVPAs the length of which is proportional to $P$. Contours levels are (\% of peak emission of $0.39\,\mathrm{Jy/beam}$): $-2$, 2, 2.83, 4, 5.65, 8, 11.31, 16, 22.63, 32, 45.25, 64, 90.51.\vspace{1.6cm}}
    \label{fig:polmap}
\end{SCfigure*}

From the Gaussian modelfits, we calculate $T_\mathrm{b}$ as:
\begin{align}
    T_\mathrm{b} = \dfrac{S_\nu c^2}{2\,k_\mathrm{B} \nu^2 \Omega}\,,\label{eq:Tb}
\end{align}
where $S_\nu$ denotes the flux density, $c$ the speed of light, $k_\mathrm{B}$ the Boltzmann constant, $\nu$ the observing frequency and $\Omega$ the component solid angle. The highest $T_\mathrm{b}$, that we calculate for component L3 (see Table~\ref{tab:3C345_modelfit} and Fig.~\ref{fig:polmap}), lies between $T_\mathrm{b,min}$ and $T_\mathrm{b,max}$ at $8.96\times10^{12}\,\mathrm{K}$, close to the maximum value. So our values of $T_\mathrm{b,min}$ and $T_\mathrm{b,max}$ seem to provide a reasonable bracketing for the highest component brightness temperature. Other studies also found a good agreement for both estimates \citep[e.g.][]{2019A&A...622A..92N}. A decline in component brightness temperature downstream of component L3 along the jet is observed, that can be explained in the framework of a jet with regions of relativistic plasma that expand adiabatically and lose energy via radiation while travelling downstream \citep{2012A&A...544A..34P}.

It is generally thought that, for incoherent synchrotron sources such as AGN, if $T_\mathrm{b}$ increases to values larger than about $10^{12}\,\mathrm{K}$, the amount of energy released due to the Inverse Compton (IC) process becomes too large to be sustainable. This `IC catastrophe' reduces $T_\mathrm{b}$ again to values of about $10^{12}\,\mathrm{K}$ on timescales of a day \citep{1969ApJ...155L..71K}. This threshold is referred to as the IC limit. \cite{1994ApJ...426...51R} argued that the equipartition brightness temperature $T_\mathrm{b,eq}$ might be a better constraint for the upper value of the brightness temperature, which is generally more of the order of $10^{11}\,\mathrm{K}$. It assumes an equipartition of the energy of particles and magnetic fields. For 3C\,345, we estimate a value very close to $10^{11}\,\mathrm{K}$ ($T_\mathrm{b,eq}=10^{10.7}$) as well, using equation (4b) in \cite{1994ApJ...426...51R} with an optically thin spectral index $\alpha=-0.2$ ($S_\nu\,\propto\,\nu^\alpha$) \citep{2018Galax...6...49L} and our single dish flux density as a proxy for the peak flux density. The equation gives an upper limit on $T_\mathrm{b,eq}$ in case we are not using the actual peak flux density of the spectrum.
We consider the source redshift $z$ and the Doppler boosting according to
\begin{align}
    T_\mathrm{b,obs} = \delta \dfrac{T_\mathrm{b,int}}{(1+z)}\,,
\end{align}
where $T_\mathrm{b,obs}$ denotes the brightness temperature in the observer's frame and $T_\mathrm{b,int}$ in the source frame. The Doppler factor is denoted by $\delta=\sqrt{1-\beta^2}(1-\beta\cos(\theta))^{-1}$, where $\beta$ is the jet bulk velocity in units of the speed of light and $\theta$ is the jet viewing angle. We take $\delta=9.1\pm{1.9}$ as reported in \cite{2017A&A...602A.104L}, calculated from variability arguments. They also constrain the Doppler factor with assumptions on the IC emission \citep{1993ApJ...407...65G}, yielding similar results. VLBI monitoring within the VLBA-BU-BLAZAR Program at 43~GHz also shows Doppler factors of the order of 10 \citep{2017ApJ...846...98J}. From these corrections we expect the theoretical value not to exceed $T_\mathrm{b}=5.7\times10^{12}\,\mathrm{K}$ in the source frame.

The visibility amplitudes imply the presence of emitting regions with observed brightness temperature in excess of this IC limit. This suggests either locally efficient injection or re-acceleration of particles in the jet to counter the inverse Compton cooling, or that the geometry of the jet creates significant changes in the Doppler factor, resulting in the sufficiently large Doppler boosting. Efficient particle re-acceleration could, for example, be achieved with turbulent plasma flowing down the jet and crossing a standing shock \citep{2014ApJ...780...87M}. Alternatively, magnetic reconnection events can efficiently accelerate particles \citep[e.g.][]{2015MNRAS.450..183S}.

Doppler boosting due to changes in the viewing angle along the jet has been investigated by \cite{1996A&A...308..395Q} for 3C\,345, who find that the position and flux variability of a component could be explained with helical motion. A similar well-pronounced case of a helical jet pattern is known, for example, in the source 1156$+$295 \citep{2004A&A...417..887H, 2011A&A...529A.113Z} or in 2136$+$141 \citep{2006ApJ...647..172S}. Variations in the jet orientation for the innermost 1~mas of the jet in 3C\,345 within about $60\,^\circ$ over 15 years also support such a scenario \citep{2013AJ....146..120L}, and the helical motion could possibly be explained by precession of the accretion disc \citep{2005A&A...431..831L}. \cite{1996A&A...308..395Q} find variability in the Doppler factor between 7 and 10.8, caused by the difference in viewing angle. This would be insufficient to explain the high $T_\mathrm{b,min}$ in our \textsl{RadioAstron} observations, where we would need $\delta>11$. \cite{2012A&A...537A..70S} have investigated the Doppler factor for different components in 3C\,345 observed between 2008 and 2010 at $43\,\mathrm{GHz}$. They find Doppler factors as high as 23 for one component. \cite{2017ApJ...846...98J} also find maximum Doppler factors of about 17. Such high Doppler boosting could readily explain the high observed brightness temperatures, however we can not identify individual components in our \textsl{RadioAstron} map with the components presented in these works.

Still, our inferred brightness temperatures easily exceed the estimated $T_\mathrm{b,eq}$ by an order of magnitude, suggesting that the jet in 3C\,345 is not in equipartition. This indicates a flaring event, that is also supported by the radio light curve and the bright polarization component we observe at $1.5\,\mathrm{mas}$ (see \ref{sec:P_str}).

More extreme values for the brightness temperature have been found in other sources observed by \textsl{RadioAstron}, for example in 3C\,273 \citep{2016ApJ...820L...9K} and BL\,Lac  \citep{2016ApJ...817...96G}. \cite{2016ApJ...820L...9K} suggest refractive substructure as a possible source of high observed $T_\mathrm{b}$, which has been investigated by \cite{2016ApJ...820L..10J}. We also test the possible effect of scattering on our estimated brightness temperature. The effect is more prominent at longer wavelengths and starts contributing to the observed signal at $18\,\mathrm{cm}$ at any baseline larger than $\sim70,000\,\mathrm{km}$ ($5.5\,D_\oplus$), if the flux density at zero spacing is $>1\,\mathrm{Jy}$ and $T_\mathrm{b}$ is in the range of the values that we also obtain here \citep[see Fig.~1 and 3 in][]{2016ApJ...820L..10J}.

\begin{SCfigure*}
    \caption{MOJAVE image of the total intensity and polarised emission of 3C\,345 at 15\,GHz. Map of the polarised intensity $P$ in colour-scale, overlaid with contours displaying the total intensity emission. The beam size is displayed on the bottom right with a resolution of $0.63\times0.46\,\mathrm{mas}$. The lines show the EVPAs the length of which is proportional to $P$. Contours levels are ($\%$ of peak emission of $3.44\,\mathrm{Jy/beam}$): $-0.031$, 0.031, 0.063, 0.125, 0.25, 0.5, 1.0, 2, 2.83, 4, 5.65, 8, 11.31, 16, 22.63, 32, 45.25, 64, 90.51.\vspace{1.3cm}}\hspace*{8pt}
    \includegraphics[width=140mm]{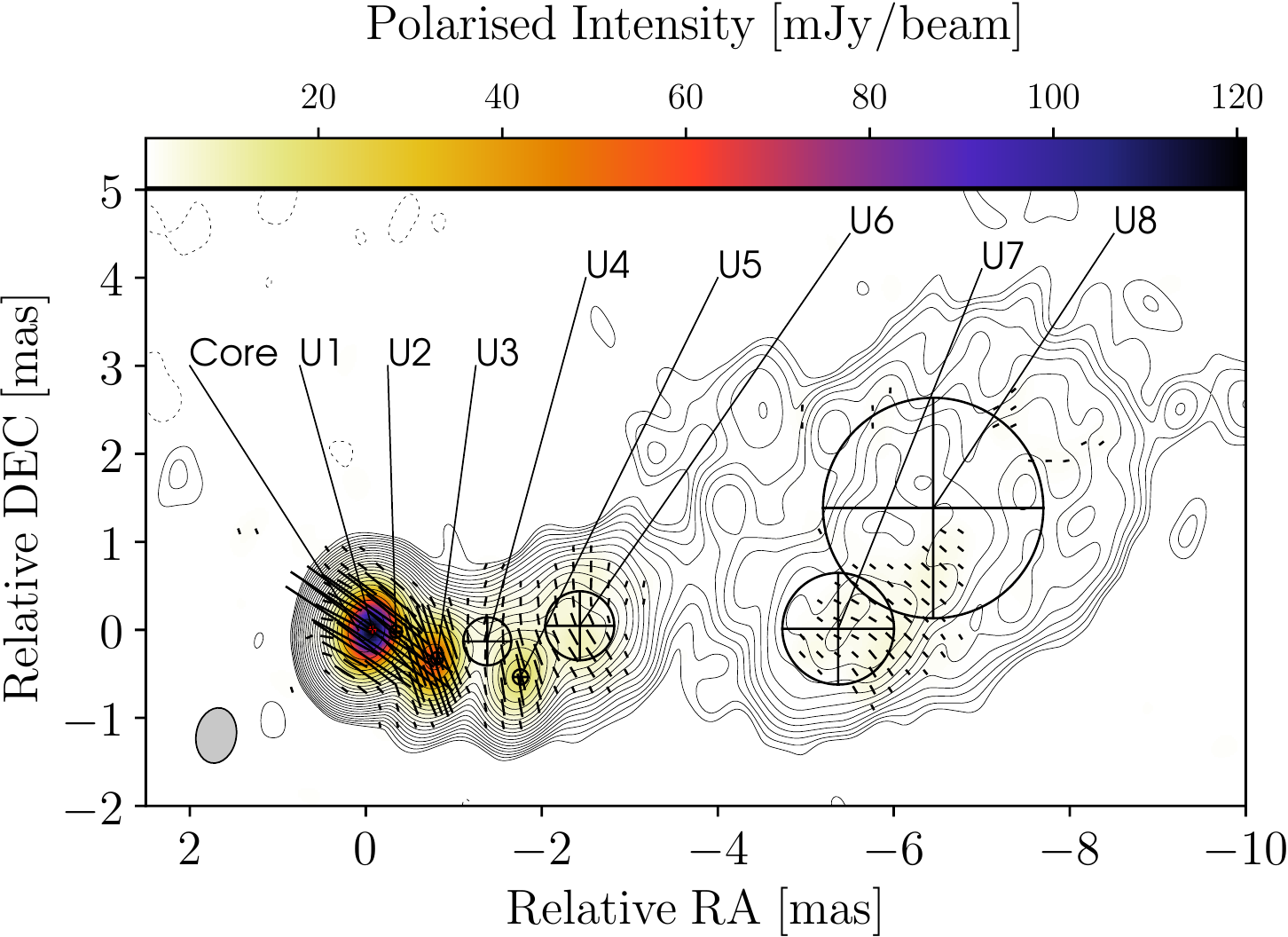}
    \label{fig:polmap_MOJ}
\end{SCfigure*}

We used Eq.~3 in \cite{2016ApJ...820L..10J} to calculate $T_\mathrm{b,min}$, which accounts for both refractive substructure and angular broadening. The former will lead to an overestimate, the latter to an underestimate of $T_\mathrm{b,min}$:
\begin{align}
    T_\mathrm{b,min} &= 1.2\times10^{12}\,\mathrm{K}\left(\dfrac{B}{10^5\,\mathrm{km}}\right)^{5/6}\left(\dfrac{F_\mathrm{B}}{20\,\mathrm{mJy}}\right)\\
    \nonumber&\times\left(\dfrac{D}{1\,\mathrm{kpc}}\right)^{1/6}\left(\dfrac{\lambda}{18\,\mathrm{cm}}\right)\left(\dfrac{\theta_\mathrm{scatt}}{300\,\mu\mathrm{as}}\right)^{-5/6}\,,
\end{align}
where $B$ denotes the baseline length and $F_B$ the measured flux density at this baseline. 
Adopting the NE2001 model \citep{2002astro.ph..7156C} for the Galactic distribution of free electrons, we estimate for the galactic coordinates of 3C\,345 ($l = 63^{\circ}.45$,  $b=40^{\circ}.95$) an angular broadening of $\theta_\mathrm{scatt}=0.28\,\mathrm{mas}$ at 18\,cm wavelength. At an assumed distance to the scattering screen of $D=1\,\mathrm{kpc}$, we estimate $T_\mathrm{b,min}=1.5\times10^{13}\,\mathrm{K}$, which is even higher than our previous estimate without considering scattering. So we conclude that refractive substructure likely does not play a role for our observations. This is not surprising as 3C\,345 lies at high galactic latitude, so there is likely not enough scattering material along the line of sight to cause significant refractive substructure.

\subsection{Comparison with brightness temperatures from MOJAVE data}\label{sec:Tb_MOJ}

We have compared the brightness temperatures obtained from the 1.6~GHz Space VLBI data with estimates obtained via the same methods using MOJAVE 15\,GHz observations. To make a reasonable comparison of the two data sets, we applied a filter on the visibilities of the MOJAVE data, so that only data that occupy the same location (within 10\,\%) in the $(u,v)$-plane as our 1.6~GHz data are used for the MOJAVE brightness temperature estimate. We present $T_\mathrm{b,min}$ as a function of $(u,v)$-radius in Fig.~\ref{fig:UV_Tb_minmax}. As for the \textsl{RadioAstron} data, we plot both the brightness temperatures obtained from the visibilities as well as those obtained from modelfits.

The range between $T_\mathrm{b,min}$ and $T_\mathrm{b,max}$ is larger for the MOJAVE data compared to our \textsl{RadioAstron} data. This is likely due to the underestimated visibility errors in the former, so $T_\mathrm{b,max}$ might not be well defined, leading to a poor determination of $T_\mathrm{b,max}$. Nevertheless, the maximum brightness temperature from modelfits to the MOJAVE data still lies between the limits provided by $T_\mathrm{b,min}$ and $T_\mathrm{b,max}$. Overall, the observed $T_\mathrm{b}$ for the Gaussian components is higher at 1.6\,GHz compared to 15\,GHz. This is expected given Eq.~\ref{eq:Tb} and the similar covered $(u,v)$-distances in both data sets. We concentrate on the comparison between the different $T_\mathrm{b,min}$ in the following. 

We see significantly higher values for $T_\mathrm{b,min}$ in the 1.6~GHz \textsl{RadioAstron} data compared to the 15~GHz MOJAVE data. That is expected, as $T_\mathrm{b}\,\propto\,\lambda^2$, and any differences in the ratio $T_\mathrm{b,min,RA}/T_\mathrm{b,min,MOJ}$ that differs from $(18\,\mathrm{cm}/2\,\mathrm{cm})^2$ as a function of $(u,v)$-distance seen in Fig.~\ref{fig:UV_Tb_minmax} can be interpreted as a spectral index that is different from zero. Indeed we observe a trend of decreasing ratios for increasing $(u,v)$-distances, which can be interpreted as a progressive change of the jet opacity from optically thin to thick with increasing baseline lengths.

%% file: Sec7_P_structure.tex
\subsection{Polarised intensity structure}\label{sec:P_str}

We observe multiple polarised components, the brightest being roughly coincident with the total intensity peak close to component L2 (see Fig.~\ref{fig:polmap}). We see more polarised structure $\sim 5\,\mathrm{mas}$ downstream of the jet. One can see an almost unpolarised core ($m=1.56\pm0.67\,\%$), where synchrotron self-absorption likely also leads to significant depolarisation. It can not be ruled out that depolarisation due to blending of different unresolved features in the observing beam also contributes to the diminished polarisation degree. An optically thick core region has been observed already between 8.1 and 15.4 GHz by MOJAVE \citep{2014AJ....147..143H}. At the total intensity peak the fractional polarisation reaches $m=5.29\pm0.51\,\%$, where at the location of component L6 it reaches $m=6.69\pm1.40\,\%$. We observe a degree of polarisation up to $60\,\%$ $\sim 5\mathrm{mas}$ downstream of the jet. This value is very close to the theoretical limit of $\sim70\,\%$. \cite{2020ApJ...893...68K} also observed up to $50\,\%$ degree of linear polarisation in the jet of 0716+714. However, the region we identify with such high fractional polarisation is situated in a low-S/N region in Stokes $I$, which drives the uncertainty of this value to be $\sim 20\,\%$. In that case we can not rule out the possibility that uncertainties in the $D$-term estimation affect the observed degree of polarisation in that region substantially.

Overall the EVPAs seem to be well aligned with the local jet direction, which was also observed at 43\,GHz by \cite{2017ApJ...850...87M}. However, in the core, the EVPAs are oriented closer to the perpendicular orientation relative to the jet. This is consistent with a possible rotation of the EVPAs by $\pi/2$ due to opacity effects \citep{1994A&A...292...33G, 2001MNRAS.320L..49G}, as the core is most likely optically thick (P\"otzl et al., in prep., from now on paper II). This would indicate a B-field closer to the perpendicular relative to the jet direction also in the core. However, \cite{2018Galax...6....5W} argues that higher optical depths of between 6 and 7 are needed for a $\pi/2$ flip in the EVPAs, which is not observed for the core region in 3C\,345.
The motion of a shock on a helical path along the jet can explain the observed bright polarised features with the EVPAs aligned with the jet direction, where the magnetic field is quenched perpendicular to the jet direction \citep{1994ApJ...437..122W}. Earlier multi-frequency, multi-epoch studies of 3C\,345 have favoured this scenario \citep{2000A&A...354...55R}. \cite{1999ApJ...521..509L} argued that shocks likely do not play a significant role in the dynamics and emission outside of the core region in 3C\,345. In this case, the EVPAs $\sim5$\,mas downstream of the core could also be explained by a large-scale helical magnetic field with a dominant toroidal component.
There is an indication for a slight gradient in EVPA direction from the inner to the outer jet region, which is however not sufficient to confirm the presence of a helical magnetic field. This will be further tested with an analysis of the Rotation Measure (paper II), as well as with higher resolution \textsl{RadioAstron} observations at $22\,\mathrm{GHz}$ made in May 2016, close to our epoch at 1.6\,GHz.

\begin{figure}
    \includegraphics[width=0.475\textwidth]{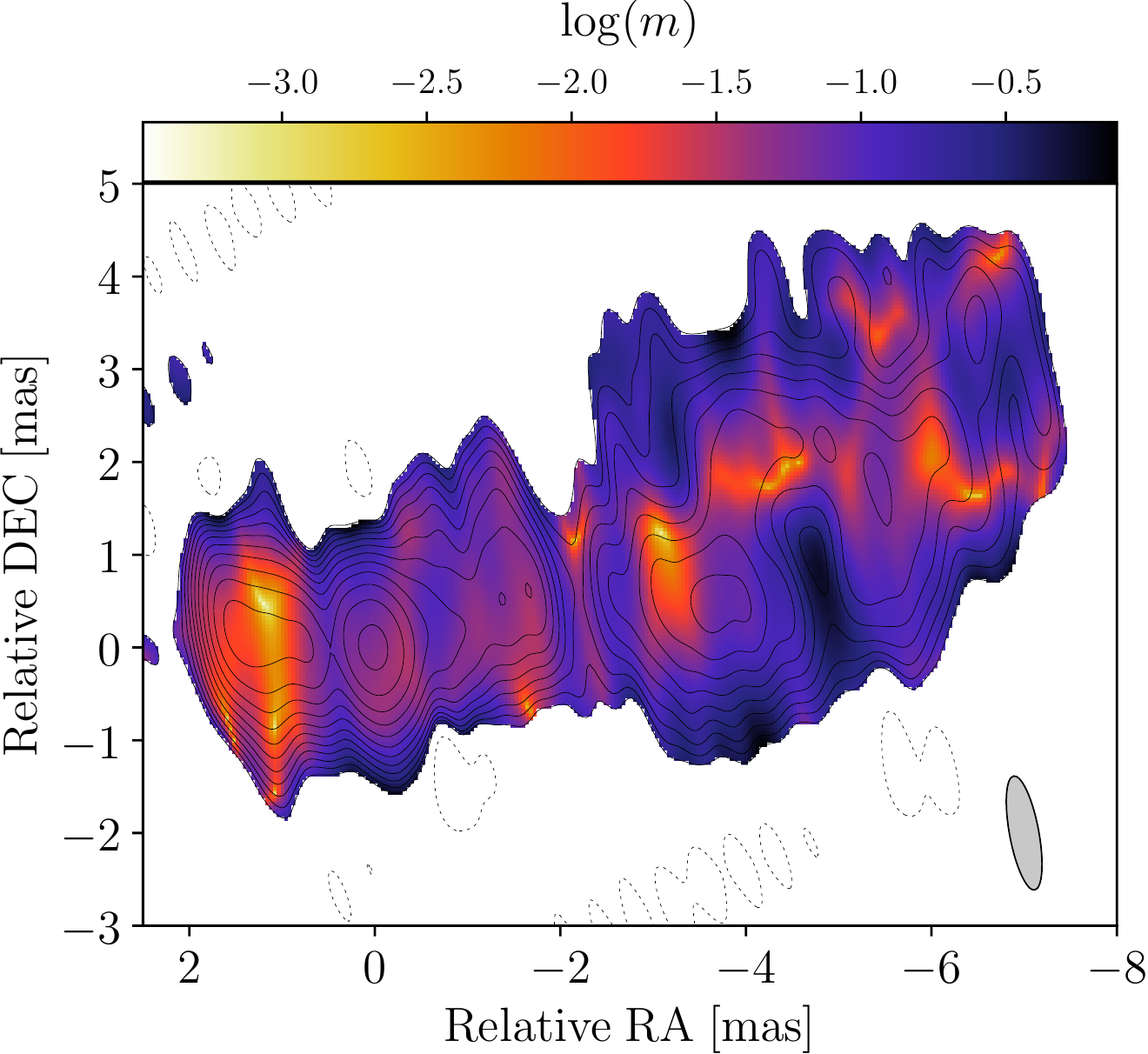}
    \caption{Same as Fig.~\ref{fig:polmap}, but displaying the logarithm of fractional polarisation ${\log}(m)$ in colour-scale.}
    \label{fig:frac_polmap}
\end{figure}

Comparing our polarisation map at $1.6\,\mathrm{GHz}$ with the MOJAVE one at $15\,\mathrm{GHz}$, we see very similar features as the ones seen $\sim 4\,\mathrm{mas}$ downstream of the jet. They may well correspond to the same features, as the jet is likely optically thin in this portion of the jet. We even see the same weakly polarised features at the northern edge of the jet. Overall the observed structure is also very similar in total intensity. The exact alignment of the two maps will be discussed in paper II.

Faraday rotation might significantly rotate the EVPAs in our observations, as their angular rotation $\Delta\chi\,\propto\,\lambda^2$, thus the Faraday rotation is stronger at lower frequencies. While we present a deeper analysis of Faraday rotation in future work using a set of data at multiple frequencies, we briefly discuss the possible magnitude of Faraday rotation.

\cite{2012AJ....144..105H} have studied the RM in many AGN jets, with observations at four frequencies between 8 and 15\,GHz. The results showed two distinct regions in 3C\,345, one with $\mathrm{RM}=156.4{\pm}72.0\,\mathrm{rad}\,\mathrm{m}^{-2}$ in the core region and another one with $\mathrm{RM}=-50.3{\pm}72.0\,\mathrm{rad}\,\mathrm{m}^{-2}$ at ${\sim}2.5\,\mathrm{mas}$ downstream of the jet. Given the measurement errors, the RM in the second region is consistent with zero, while for the core-region one could expect a rotation of up to $\sim64^\circ$, which would significantly change the EVPAs in the core. \cite{2017MNRAS.467.2648M} also studied the RM at four frequencies around 1.6\,GHz in six AGN, one of which was 3C\,345. They found RM in the range of $-30\,\mathrm{rad}\,\mathrm{m}^{-2}<RM<30\,\mathrm{rad}\,\mathrm{m}^{-2}$, and report a statistically significant RM gradient transverse to the jet direction. This supports the presence of a toroidal magnetic field that may be part of a helical one. However, the difference in beam size compared to our \textsl{RadioAstron} observations is about a factor of 20 in the east-west and a factor 10 in the north-south direction, and our observations only focus on the innermost $\sim10\,\mathrm{mas}$ of the jet.

%% file: Sec8_Summary.tex
\section{Summary}

The main conclusions of the paper are summarised in the following:

\begin{itemize}
    \item We present Space VLBI images obtained with the \textsl{RadioAstron} mission in both total and linearly polarised intensity of the FSRQ 3C\,345 at 1.6\,GHz with an angular resolution of $\sim300\,\mu$as. Several compact components that were not identifiable with ground-only VLBI arrays at the same frequency are resolved in our \textsl{RadioAstron} observations and the Space VLBI image reveals the complex, visibly curved inner jet structure.
    \item We identify several linearly polarised components, with an almost completely depolarised core, a high polarisation peak coincident with the total intensity peak with $\sim6\,\%$ degree of linear polarisation, as well as more distinct components downstream of the jet. The EVPAs in those components align well with the jet direction, indicating a magnetic field perpendicular to the jet flow. The nature of these components is likely to be related to shocks propagating along a helical path of the jet. Another possibility includes a large-scale helical magnetic field with a dominant toroidal component.
    \item We compare several estimates of the brightness temperature $T_\mathrm{b}$ for the \textsl{RadioAstron} data. We infer a minimum observed brightness temperature of $T_\mathrm{b,min,obs}=6.5\times10^{12}\,\mathrm{K}$ and a minimum intrinsic brightness temperature $T_\mathrm{b,min,int}=1.1\times10^{12}\,\mathrm{K}$. The latter is in slight excess of the IC limit, and an order of magnitude larger than the equipartition brightness temperature limit, suggesting that 3C\,345 is not in equipartition between particle and magnetic field energy during our observations. The most likely explanations of this excess are either a variable Doppler factor ($\delta>11$) due to changes in the jet geometry along the flow or locally efficient particle re-acceleration. We investigated the effect of refractive substructure due to the galactic ISM and conclude that it does not dominate our estimate. We also confirm that the range given by $T_\mathrm{b,min}$ and $T_\mathrm{b,max}$ accurately brackets the actual $T_\mathrm{b}$ as measured from fitting the data with circular Gaussian components.
\end{itemize}

These conclusions will be further tested with an analysis of the \textsl{RadioAstron} data presented here in conjunction with a multiwavelength VLBI dataset in P{\"o}tzl et al.~(in prep.).

%% file: MAIN.bbl
\begin{thebibliography}{76}
\expandafter\ifx\csname natexlab\endcsname\relax\def\natexlab#1{#1}\fi

\bibitem[{{Angelakis} {et~al.}(2019){Angelakis}, {Fuhrmann}, {Myserlis},
  {Zensus}, {Nestoras}, {Karamanavis}, {Marchili}, {Krichbaum}, {Kraus}, \&
  {Rachen}}]{2019A&A...626A..60A}
{Angelakis}, E., {Fuhrmann}, L., {Myserlis}, I., {et~al.} 2019, \aap, 626, A60

\bibitem[{{Blandford} \& {Payne}(1982)}]{1982MNRAS.199..883B}
{Blandford}, R.~D. \& {Payne}, D.~G. 1982, \mnras, 199, 883

\bibitem[{{Blandford} \& {Znajek}(1977)}]{1977MNRAS.179..433B}
{Blandford}, R.~D. \& {Znajek}, R.~L. 1977, \mnras, 179, 433

\bibitem[{{Bruni} {et~al.}(2016){Bruni}, {Anderson}, {Alef}, {Rottmann},
  {Lobanov}, \& {Zensus}}]{2016Galax...4...55B}
{Bruni}, G., {Anderson}, J., {Alef}, W., {et~al.} 2016, Galaxies, 4, 55

\bibitem[{{Bruni} {et~al.}(2017){Bruni}, {G{\'o}mez}, {Casadio}, {Lobanov},
  {Kovalev}, {Sokolovsky}, {Lisakov}, {Bach}, {Marscher}, {Jorstad},
  {Anderson}, {Krichbaum}, {Savolainen}, {Vega-Garc{\'\i}a}, {Fuentes},
  {Zensus}, {Alberdi}, {Lee}, {Lu}, {P{\'e}rez-Torres}, \&
  {Ros}}]{2017A&A...604A.111B}
{Bruni}, G., {G{\'o}mez}, J.~L., {Casadio}, C., {et~al.} 2017, \aap, 604, A111

\bibitem[{{Bruni} {et~al.}(2020){Bruni}, {Savolainen}, {G{\'o}mez}, {Lobanov},
  {Kovalev}, {RadioAstron AGN Imaging Team}, \& {KSP
  Team}}]{2020AdSpR..65..712B}
{Bruni}, G., {Savolainen}, T., {G{\'o}mez}, J.~L., {et~al.} 2020, Advances in
  Space Research, 65, 712

\bibitem[{{Cordes} \& {Lazio}(2002)}]{2002astro.ph..7156C}
{Cordes}, J.~M. \& {Lazio}, T.~J.~W. 2002, arXiv e-prints
  [\eprint{astro-ph/0207156}]

\bibitem[{{Deller} {et~al.}(2011){Deller}, {Brisken}, {Phillips}, {Morgan},
  {Alef}, {Cappallo}, {Middelberg}, {Romney}, {Rottmann}, {Tingay}, \&
  {Wayth}}]{2011PASP..123..275D}
{Deller}, A.~T., {Brisken}, W.~F., {Phillips}, C.~J., {et~al.} 2011, \pasp,
  123, 275

\bibitem[{{Event Horizon Telescope Collaboration} {et~al.}(2019){Event Horizon
  Telescope Collaboration}, {Akiyama}, {Alberdi}, {Alef}, {Asada}, {Azulay},
  {Baczko}, {Ball}, {Balokovi{\'c}}, {Barrett}, \&
  et~al.}]{2019ApJ...875L...5E}
{Event Horizon Telescope Collaboration}, {Akiyama}, K., {Alberdi}, A., {et~al.}
  2019, \apjl, 875, L5

\bibitem[{{Gabuzda} \& {G{\'o}mez}(2001)}]{2001MNRAS.320L..49G}
{Gabuzda}, D.~C. \& {G{\'o}mez}, J.~L. 2001, \mnras, 320, L49

\bibitem[{{Ghisellini} {et~al.}(1993){Ghisellini}, {Padovani}, {Celotti}, \&
  {Maraschi}}]{1993ApJ...407...65G}
{Ghisellini}, G., {Padovani}, P., {Celotti}, A., \& {Maraschi}, L. 1993, \apj,
  407, 65

\bibitem[{{Giovannini} {et~al.}(2018){Giovannini}, {Savolainen}, {Orienti},
  {Nakamura}, {Nagai}, {Kino}, {Giroletti}, {Hada}, {Bruni}, {Kovalev},
  {Anderson}, {D'Ammando}, {Hodgson}, {Honma}, {Krichbaum}, {Lee}, {Lico},
  {Lisakov}, {Lobanov}, {Petrov}, {Sohn}, {Sokolovsky}, {Voitsik}, {Zensus}, \&
  {Tingay}}]{2018NatAs...2..472G}
{Giovannini}, G., {Savolainen}, T., {Orienti}, M., {et~al.} 2018, Nature
  Astronomy, 2, 472

\bibitem[{{Gomez} {et~al.}(1994){Gomez}, {Alberdi}, {Marcaide}, {Marscher}, \&
  {Travis}}]{1994A&A...292...33G}
{Gomez}, J.~L., {Alberdi}, A., {Marcaide}, J.~M., {Marscher}, A.~P., \&
  {Travis}, J.~P. 1994, \aap, 292, 33

\bibitem[{{G{\'o}mez} {et~al.}(2016){G{\'o}mez}, {Lobanov}, {Bruni}, {Kovalev},
  {Marscher}, {Jorstad}, {Mizuno}, {Bach}, {Sokolovsky}, {Anderson}, {Galindo},
  {Kardashev}, \& {Lisakov}}]{2016ApJ...817...96G}
{G{\'o}mez}, J.~L., {Lobanov}, A.~P., {Bruni}, G., {et~al.} 2016, \apj, 817, 96

\bibitem[{{G{\'o}mez} {et~al.}(2011){G{\'o}mez}, {Roca-Sogorb}, {Agudo},
  {Marscher}, \& {Jorstad}}]{2011ApJ...733...11G}
{G{\'o}mez}, J.~L., {Roca-Sogorb}, M., {Agudo}, I., {Marscher}, A.~P., \&
  {Jorstad}, S.~G. 2011, \apj, 733, 11

\bibitem[{{Greisen}(2003)}]{2003ASSL..285..109G}
{Greisen}, E.~W. 2003, Astrophysics and Space Science Library, Vol. 285, {AIPS,
  the VLA, and the VLBA}, ed. A.~{Heck}, 109

\bibitem[{{Gu} {et~al.}(2001){Gu}, {Cao}, \& {Jiang}}]{2001MNRAS.327.1111G}
{Gu}, M., {Cao}, X., \& {Jiang}, D.~R. 2001, \mnras, 327, 1111

\bibitem[{{Hong} {et~al.}(2004){Hong}, {Jiang}, {Gurvits}, {Garrett},
  {Garrington}, {Schilizzi}, {Nan}, {Hirabayashi}, {Wang}, \&
  {Nicolson}}]{2004A&A...417..887H}
{Hong}, X.~Y., {Jiang}, D.~R., {Gurvits}, L.~I., {et~al.} 2004, \aap, 417, 887

\bibitem[{{Hovatta} {et~al.}(2014){Hovatta}, {Aller}, {Aller}, {Clausen-Brown},
  {Homan}, {Kovalev}, {Lister}, {Pushkarev}, \&
  {Savolainen}}]{2014AJ....147..143H}
{Hovatta}, T., {Aller}, M.~F., {Aller}, H.~D., {et~al.} 2014, \aj, 147, 143

\bibitem[{{Hovatta} {et~al.}(2012){Hovatta}, {Lister}, {Aller}, {Aller},
  {Homan}, {Kovalev}, {Pushkarev}, \& {Savolainen}}]{2012AJ....144..105H}
{Hovatta}, T., {Lister}, M.~L., {Aller}, M.~F., {et~al.} 2012, \aj, 144, 105

\bibitem[{{Johnson} {et~al.}(2016){Johnson}, {Kovalev}, {Gwinn}, {Gurvits},
  {Narayan}, {Macquart}, {Jauncey}, {Voitsik}, {Anderson}, {Sokolovsky}, \&
  {Lisakov}}]{2016ApJ...820L..10J}
{Johnson}, M.~D., {Kovalev}, Y.~Y., {Gwinn}, C.~R., {et~al.} 2016, \apjl, 820,
  L10

\bibitem[{{Jorstad} {et~al.}(2017){Jorstad}, {Marscher}, {Morozova},
  {Troitsky}, {Agudo}, {Casadio}, {Foord}, {G{\'o}mez}, {MacDonald}, {Molina},
  {L{\"a}hteenm{\"a}ki}, {Tammi}, \& {Tornikoski}}]{2017ApJ...846...98J}
{Jorstad}, S.~G., {Marscher}, A.~P., {Morozova}, D.~A., {et~al.} 2017, \apj,
  846, 98

\bibitem[{{Kardashev} {et~al.}(2013){Kardashev}, {Khartov}, {Abramov},
  {Avdeev}, {Alakoz}, {Aleksandrov}, {Ananthakrishnan}, {Andreyanov},
  {Andrianov}, {Antonov}, {Artyukhov}, {Arkhipov}, {Baan}, {Babakin},
  {Babyshkin}, {Bartel'}, {Belousov}, {Belyaev}, {Berulis}, {Burke},
  {Biryukov}, {Bubnov}, {Burgin}, {Busca}, {Bykadorov}, {Bychkova},
  {Vasil'kov}, {Wellington}, {Vinogradov}, {Wietfeldt}, {Voitsik},
  {Gvamichava}, {Girin}, {Gurvits}, {Dagkesamanskii}, {D'Addario},
  {Giovannini}, {Jauncey}, {Dewdney}, {D'yakov}, {Zharov}, {Zhuravlev},
  {Zaslavskii}, {Zakhvatkin}, {Zinov'ev}, {Ilinen}, {Ipatov}, {Kanevskii},
  {Knorin}, {Casse}, {Kellermann}, {Kovalev}, {Kovalev}, {Kovalenko}, {Kogan},
  {Komaev}, {Konovalenko}, {Kopelyanskii}, {Korneev}, {Kostenko}, {Kotik},
  {Kreisman}, {Kukushkin}, {Kulishenko}, {Cooper}, {Kut'kin}, {Cannon},
  {Larionov}, {Lisakov}, {Litvinenko}, {Likhachev}, {Likhacheva}, {Lobanov},
  {Logvinenko}, {Langston}, {McCracken}, {Medvedev}, {Melekhin}, {Menderov},
  {Murphy}, {Mizyakina}, {Mozgovoi}, {Nikolaev}, {Novikov}, {Novikov},
  {Oreshko}, {Pavlenko}, {Pashchenko}, {Ponomarev}, {Popov}, {Pravin-Kumar},
  {Preston}, {Pyshnov}, {Rakhimov}, {Rozhkov}, {Romney}, {Rocha}, {Rudakov},
  {R{\"a}is{\"a}nen}, {Sazankov}, {Sakharov}, {Semenov}, {Serebrennikov},
  {Schilizzi}, {Skulachev}, {Slysh}, {Smirnov}, {Smith}, {Soglasnov},
  {Sokolovskii}, {Sondaar}, {Stepan'yants}, {Turygin}, {Turygin}, {Tuchin},
  {Urpo}, {Fedorchuk}, {Finkel'shtein}, {Fomalont}, {Fejes}, {Fomina},
  {Khapin}, {Tsarevskii}, {Zensus}, {Chuprikov}, {Shatskaya}, {Shapirovskaya},
  {Sheikhet}, {Shirshakov}, {Schmidt}, {Shnyreva}, {Shpilevskii}, {Ekers}, \&
  {Yakimov}}]{2013ARep...57..153K}
{Kardashev}, N.~S., {Khartov}, V.~V., {Abramov}, V.~V., {et~al.} 2013,
  Astronomy Reports, 57, 153

\bibitem[{{Kellermann} \& {Pauliny-Toth}(1969)}]{1969ApJ...155L..71K}
{Kellermann}, K.~I. \& {Pauliny-Toth}, I.~I.~K. 1969, \apjl, 155, L71

\bibitem[{{Klare} {et~al.}(2005){Klare}, {Zensus}, {Lobanov}, {Ros},
  {Krichbaum}, \& {Witzel}}]{2005ASPC..340...40K}
{Klare}, J., {Zensus}, J.~A., {Lobanov}, A.~P., {et~al.} 2005, in Astronomical
  Society of the Pacific Conference Series, Vol. 340, Future Directions in High
  Resolution Astronomy, ed. J.~{Romney} \& M.~{Reid}, 40

\bibitem[{{Klare} {et~al.}(2000){Klare}, {Zensus}, {Ros}, \&
  {Lobanov}}]{2000aprs.conf...21K}
{Klare}, J., {Zensus}, J.~A., {Ros}, E., \& {Lobanov}, A.~P. 2000, in
  Astrophysical Phenomena Revealed by Space VLBI, ed. H.~{Hirabayashi}, P.~G.
  {Edwards}, \& D.~W. {Murphy}, 21--24

\bibitem[{{Koay} {et~al.}(2019){Koay}, {Jauncey}, {Hovatta}, {Kiehlmann},
  {Bignall}, {Max-Moerbeck}, {Pearson}, {Readhead}, {Reeves}, {Reynolds}, \&
  {Vedantham}}]{2019MNRAS.489.5365K}
{Koay}, J.~Y., {Jauncey}, D.~L., {Hovatta}, T., {et~al.} 2019, \mnras, 489,
  5365

\bibitem[{{Koay} {et~al.}(2018){Koay}, {Macquart}, {Jauncey}, {Pursimo},
  {Giroletti}, {Bignall}, {Lovell}, {Rickett}, {Kedziora-Chudczer}, {Ojha}, \&
  {Reynolds}}]{2018MNRAS.474.4396K}
{Koay}, J.~Y., {Macquart}, J.~P., {Jauncey}, D.~L., {et~al.} 2018, \mnras, 474,
  4396

\bibitem[{{Kovalev} {et~al.}(2014){Kovalev}, {Vasil'kov}, {Popov}, {Soglasnov},
  {Voitsik}, {Lisakov}, {Kut'kin}, {Nikolaev}, {Nizhel'skii}, {Zhekanis}, \&
  {Tsybulev}}]{2014CosRe..52..393K}
{Kovalev}, Y.~A., {Vasil'kov}, V.~I., {Popov}, M.~V., {et~al.} 2014, Cosmic
  Research, 52, 393

\bibitem[{{Kovalev} {et~al.}(2016){Kovalev}, {Kardashev}, {Kellermann},
  {Lobanov}, {Johnson}, {Gurvits}, {Voitsik}, {Zensus}, {Anderson}, {Bach},
  {Jauncey}, {Ghigo}, {Ghosh}, {Kraus}, {Kovalev}, {Lisakov}, {Petrov},
  {Romney}, {Salter}, \& {Sokolovsky}}]{2016ApJ...820L...9K}
{Kovalev}, Y.~Y., {Kardashev}, N.~S., {Kellermann}, K.~I., {et~al.} 2016,
  \apjl, 820, L9

\bibitem[{{Kovalev} {et~al.}(2020{\natexlab{a}}){Kovalev}, {Kardashev},
  {Sokolovsky}, {Voitsik}, {An}, {Anderson}, {Andrianov}, {Avdeev}, {Bartel},
  {Bignall}, {Burgin}, {Edwards}, {Ellingsen}, {Frey}, {Garc{\'\i}a-Mir{\'o}},
  {Gawro{\'n}ski}, {Ghigo}, {Ghosh}, {Giovannini}, {Girin}, {Giroletti},
  {Gurvits}, {Jauncey}, {Horiuchi}, {Ivanov}, {Kharinov}, {Koay}, {Kostenko},
  {Kovalenko}, {Kovalev}, {Kravchenko}, {Kunert-Bajraszewska}, {Kutkin},
  {Likhachev}, {Lisakov}, {Litovchenko}, {McCallum}, {Melis}, {Melnikov},
  {Migoni}, {Nair}, {Pashchenko}, {Phillips}, {Polatidis}, {Pushkarev},
  {Quick}, {Rakhimov}, {Reynolds}, {Rizzo}, {Rudnitskiy}, {Savolainen},
  {Shakhvorostova}, {Shatskaya}, {Shen}, {Shchurov}, {Vermeulen}, {de Vicente},
  {Wolak}, {Zensus}, \& {Zuga}}]{2020AdSpR..65..705K}
{Kovalev}, Y.~Y., {Kardashev}, N.~S., {Sokolovsky}, K.~V., {et~al.}
  2020{\natexlab{a}}, Advances in Space Research, 65, 705

\bibitem[{{Kovalev} {et~al.}(2005){Kovalev}, {Kellermann}, {Lister}, {Homan},
  {Vermeulen}, {Cohen}, {Ros}, {Kadler}, {Lobanov}, {Zensus}, {Kardashev},
  {Gurvits}, {Aller}, \& {Aller}}]{2005AJ....130.2473K}
{Kovalev}, Y.~Y., {Kellermann}, K.~I., {Lister}, M.~L., {et~al.} 2005, \aj,
  130, 2473

\bibitem[{{Kovalev} {et~al.}(2020{\natexlab{b}}){Kovalev}, {Pushkarev},
  {Nokhrina}, {Plavin}, {Beskin}, {Chernoglazov}, {Lister}, \&
  {Savolainen}}]{2020MNRAS.495.3576K}
{Kovalev}, Y.~Y., {Pushkarev}, A.~B., {Nokhrina}, E.~E., {et~al.}
  2020{\natexlab{b}}, \mnras, 495, 3576

\bibitem[{{Kravchenko} {et~al.}(2020){Kravchenko}, {G{\'o}mez}, {Kovalev},
  {Lobanov}, {Savolainen}, {Bruni}, {Fuentes}, {Anderson}, {Jorstad},
  {Marscher}, {Tornikoski}, {L{\"a}hteenm{\"a}ki}, \&
  {Lisakov}}]{2020ApJ...893...68K}
{Kravchenko}, E.~V., {G{\'o}mez}, J.~L., {Kovalev}, Y.~Y., {et~al.} 2020, \apj,
  893, 68

\bibitem[{{Kutkin} {et~al.}(2018){Kutkin}, {Pashchenko}, {Lisakov}, {Voytsik},
  {Sokolovsky}, {Kovalev}, {Lobanov}, {Ipatov}, {Aller}, {Aller},
  {Lahteenmaki}, {Tornikoski}, \& {Gurvits}}]{2018MNRAS.475.4994K}
{Kutkin}, A.~M., {Pashchenko}, I.~N., {Lisakov}, M.~M., {et~al.} 2018, \mnras,
  475, 4994

\bibitem[{{Liodakis} {et~al.}(2017){Liodakis}, {Zezas}, {Angelakis}, {Hovatta},
  \& {Pavlidou}}]{2017A&A...602A.104L}
{Liodakis}, I., {Zezas}, A., {Angelakis}, E., {Hovatta}, T., \& {Pavlidou}, V.
  2017, \aap, 602, A104

\bibitem[{{Lipunov} {et~al.}(2010){Lipunov}, {Kornilov}, {Gorbovskoy},
  {Shatskij}, {Kuvshinov}, {Tyurina}, {Belinski}, {Krylov}, {Balanutsa},
  {Chazov}, {Kuznetsov}, {Kortunov}, {Sankovich}, {Tlatov}, {Parkhomenko},
  {Krushinsky}, {Zalozhnyh}, {Popov}, {Kopytova}, {Ivanov}, {Yazev}, \&
  {Yurkov}}]{2010AdAst2010E..30L}
{Lipunov}, V., {Kornilov}, V., {Gorbovskoy}, E., {et~al.} 2010, Advances in
  Astronomy, 2010, 349171

\bibitem[{{Lister} {et~al.}(2018){Lister}, {Aller}, {Aller}, {Hodge}, {Homan},
  {Kovalev}, {Pushkarev}, \& {Savolainen}}]{2018ApJS..234...12L}
{Lister}, M.~L., {Aller}, M.~F., {Aller}, H.~D., {et~al.} 2018, \apjs, 234, 12

\bibitem[{{Lister} {et~al.}(2013){Lister}, {Aller}, {Aller}, {Homan},
  {Kellermann}, {Kovalev}, {Pushkarev}, {Richards}, {Ros}, \&
  {Savolainen}}]{2013AJ....146..120L}
{Lister}, M.~L., {Aller}, M.~F., {Aller}, H.~D., {et~al.} 2013, \aj, 146, 120

\bibitem[{{Lister} {et~al.}(2019){Lister}, {Homan}, {Hovatta}, {Kellermann},
  {Kiehlmann}, {Kovalev}, {Max-Moerbeck}, {Pushkarev}, {Readhead}, {Ros}, \&
  {Savolainen}}]{2019ApJ...874...43L}
{Lister}, M.~L., {Homan}, D.~C., {Hovatta}, T., {et~al.} 2019, \apj, 874, 43

\bibitem[{{Liu} {et~al.}(2018){Liu}, {Bignall}, {Krichbaum}, {Liu}, {Kraus},
  {Kovalev}, {Sokolovsky}, {Angelakis}, \& {Zensus}}]{2018Galax...6...49L}
{Liu}, J., {Bignall}, H., {Krichbaum}, T., {et~al.} 2018, Galaxies, 6, 49

\bibitem[{{Lobanov}(2015)}]{2015A&A...574A..84L}
{Lobanov}, A. 2015, \aap, 574, A84

\bibitem[{{Lobanov}(2005)}]{2005astro.ph..3225L}
{Lobanov}, A.~P. 2005, arXiv e-prints [\eprint{astro-ph/0503225}]

\bibitem[{{Lobanov} {et~al.}(2015){Lobanov}, {G{\'o}mez}, {Bruni}, {Kovalev},
  {Anderson}, {Bach}, {Kraus}, {Zensus}, {Lisakov}, {Sokolovsky}, \&
  {Voytsik}}]{2015A&A...583A.100L}
{Lobanov}, A.~P., {G{\'o}mez}, J.~L., {Bruni}, G., {et~al.} 2015, \aap, 583,
  A100

\bibitem[{{Lobanov} \& {Roland}(2005)}]{2005A&A...431..831L}
{Lobanov}, A.~P. \& {Roland}, J. 2005, \aap, 431, 831

\bibitem[{{Lobanov} \& {Zensus}(1999)}]{1999ApJ...521..509L}
{Lobanov}, A.~P. \& {Zensus}, J.~A. 1999, \apj, 521, 509

\bibitem[{{Lovell} {et~al.}(2008){Lovell}, {Rickett}, {Macquart}, {Jauncey},
  {Bignall}, {Kedziora-Chudczer}, {Ojha}, {Pursimo}, {Dutka}, {Senkbeil}, \&
  {Shabala}}]{2008ApJ...689..108L}
{Lovell}, J.~E.~J., {Rickett}, B.~J., {Macquart}, J.~P., {et~al.} 2008, \apj,
  689, 108

\bibitem[{{MacDonald} {et~al.}(2017){MacDonald}, {Jorstad}, \&
  {Marscher}}]{2017ApJ...850...87M}
{MacDonald}, N.~R., {Jorstad}, S.~G., \& {Marscher}, A.~P. 2017, \apj, 850, 87

\bibitem[{{Marscher}(2014)}]{2014ApJ...780...87M}
{Marscher}, A.~P. 2014, \apj, 780, 87

\bibitem[{{Marziani} {et~al.}(1996){Marziani}, {Sulentic}, {Dultzin-Hacyan},
  {Calvani}, \& {Moles}}]{1996ApJS..104...37M}
{Marziani}, P., {Sulentic}, J.~W., {Dultzin-Hacyan}, D., {Calvani}, M., \&
  {Moles}, M. 1996, \apjs, 104, 37

\bibitem[{{Meier} {et~al.}(2001){Meier}, {Koide}, \&
  {Uchida}}]{2001Sci...291...84M}
{Meier}, D.~L., {Koide}, S., \& {Uchida}, Y. 2001, Science, 291, 84

\bibitem[{{Motter} \& {Gabuzda}(2017)}]{2017MNRAS.467.2648M}
{Motter}, J.~C. \& {Gabuzda}, D.~C. 2017, \mnras, 467, 2648

\bibitem[{{Nair} {et~al.}(2019){Nair}, {Lobanov}, {Krichbaum}, {Ros}, {Zensus},
  {Kovalev}, {Lee}, {Mertens}, {Hagiwara}, {Bremer}, {Lindqvist}, \& {de
  Vicente}}]{2019A&A...622A..92N}
{Nair}, D.~G., {Lobanov}, A.~P., {Krichbaum}, T.~P., {et~al.} 2019, \aap, 622,
  A92

\bibitem[{{Pashchenko} {et~al.}(2015){Pashchenko}, {Kovalev}, \&
  {Voitsik}}]{2015CosRe..53..199P}
{Pashchenko}, I.~N., {Kovalev}, Y.~Y., \& {Voitsik}, P.~A. 2015, Cosmic
  Research, 53, 199

\bibitem[{{Pilipenko} {et~al.}(2018){Pilipenko}, {Kovalev}, {Andrianov},
  {Bach}, {Buttaccio}, {Cassaro}, {Cim{\`o}}, {Edwards}, {Gawro{\'n}ski},
  {Gurvits}, {Hovatta}, {Jauncey}, {Johnson}, {Kovalev}, {Kutkin}, {Lisakov},
  {Melnikov}, {Orlati}, {Rudnitskiy}, {Sokolovsky}, {Stanghellini}, {de
  Vicente}, {Voitsik}, {Wolak}, \& {Zhekanis}}]{2018MNRAS.474.3523P}
{Pilipenko}, S.~V., {Kovalev}, Y.~Y., {Andrianov}, A.~S., {et~al.} 2018,
  \mnras, 474, 3523

\bibitem[{{Planck Collaboration} {et~al.}(2014){Planck Collaboration}, {Ade},
  {Aghanim}, {Armitage-Caplan}, {Arnaud}, {Ashdown}, {Atrio-Barand ela},
  {Aumont}, {Baccigalupi}, {Banday}, \& et~al.}]{2014A&A...571A..16P}
{Planck Collaboration}, {Ade}, P.~A.~R., {Aghanim}, N., {et~al.} 2014, \aap,
  571, A16

\bibitem[{{Pushkarev} \& {Kovalev}(2012)}]{2012A&A...544A..34P}
{Pushkarev}, A.~B. \& {Kovalev}, Y.~Y. 2012, \aap, 544, A34

\bibitem[{{Pushkarev} {et~al.}(2009){Pushkarev}, {Kovalev}, {Lister}, \&
  {Savolainen}}]{2009A&A...507L..33P}
{Pushkarev}, A.~B., {Kovalev}, Y.~Y., {Lister}, M.~L., \& {Savolainen}, T.
  2009, \aap, 507, L33

\bibitem[{{Qian} {et~al.}(1996){Qian}, {Krichbaum}, {Zensus}, {Steffen}, \&
  {Witzel}}]{1996A&A...308..395Q}
{Qian}, S.~J., {Krichbaum}, T.~P., {Zensus}, J.~A., {Steffen}, W., \& {Witzel},
  A. 1996, \aap, 308, 395

\bibitem[{{Readhead}(1994)}]{1994ApJ...426...51R}
{Readhead}, A. C.~S. 1994, \apj, 426, 51

\bibitem[{{Richards} {et~al.}(2014){Richards}, {Hovatta}, {Max-Moerbeck},
  {Pavlidou}, {Pearson}, \& {Readhead}}]{2014MNRAS.438.3058R}
{Richards}, J.~L., {Hovatta}, T., {Max-Moerbeck}, W., {et~al.} 2014, \mnras,
  438, 3058

\bibitem[{{Rickett} {et~al.}(2006){Rickett}, {Lazio}, \&
  {Ghigo}}]{2006ApJS..165..439R}
{Rickett}, B.~J., {Lazio}, T.~J.~W., \& {Ghigo}, F.~D. 2006, \apjs, 165, 439

\bibitem[{{Ros} {et~al.}(2000){Ros}, {Zensus}, \&
  {Lobanov}}]{2000A&A...354...55R}
{Ros}, E., {Zensus}, J.~A., \& {Lobanov}, A.~P. 2000, \aap, 354, 55

\bibitem[{{Sambruna} {et~al.}(2004){Sambruna}, {Gambill}, {Maraschi},
  {Tavecchio}, {Cerutti}, {Cheung}, {Urry}, \& {Chartas}}]{2004ApJ...608..698S}
{Sambruna}, R.~M., {Gambill}, J.~K., {Maraschi}, L., {et~al.} 2004, \apj, 608,
  698

\bibitem[{{Savolainen} {et~al.}(2006){Savolainen}, {Wiik}, {Valtaoja},
  {Kadler}, {Ros}, {Tornikoski}, {Aller}, \& {Aller}}]{2006ApJ...647..172S}
{Savolainen}, T., {Wiik}, K., {Valtaoja}, E., {et~al.} 2006, \apj, 647, 172

\bibitem[{{Schinzel} {et~al.}(2012){Schinzel}, {Lobanov}, {Taylor}, {Jorstad},
  {Marscher}, \& {Zensus}}]{2012A&A...537A..70S}
{Schinzel}, F.~K., {Lobanov}, A.~P., {Taylor}, G.~B., {et~al.} 2012, \aap, 537,
  A70

\bibitem[{{Shen} {et~al.}(2011){Shen}, {Richards}, {Strauss}, {Hall},
  {Schneider}, {Snedden}, {Bizyaev}, {Brewington}, {Malanushenko},
  {Malanushenko}, {Oravetz}, {Pan}, \& {Simmons}}]{2011ApJS..194...45S}
{Shen}, Y., {Richards}, G.~T., {Strauss}, M.~A., {et~al.} 2011, \apjs, 194, 45

\bibitem[{{Shepherd}(1997)}]{1997ASPC..125...77S}
{Shepherd}, M.~C. 1997, in Astronomical Society of the Pacific Conference
  Series, Vol. 125, Astronomical Data Analysis Software and Systems VI, ed.
  G.~{Hunt} \& H.~{Payne}, 77

\bibitem[{{Sironi} {et~al.}(2015){Sironi}, {Petropoulou}, \&
  {Giannios}}]{2015MNRAS.450..183S}
{Sironi}, L., {Petropoulou}, M., \& {Giannios}, D. 2015, \mnras, 450, 183

\bibitem[{{Sokolovsky} {et~al.}(2011){Sokolovsky}, {Kovalev}, {Pushkarev}, \&
  {Lobanov}}]{2011A&A...532A..38S}
{Sokolovsky}, K.~V., {Kovalev}, Y.~Y., {Pushkarev}, A.~B., \& {Lobanov}, A.~P.
  2011, \aap, 532, A38

\bibitem[{{Wardle}(2018)}]{2018Galax...6....5W}
{Wardle}, J. 2018, Galaxies, 6, 5

\bibitem[{{Wardle} {et~al.}(1994){Wardle}, {Cawthorne}, {Roberts}, \&
  {Brown}}]{1994ApJ...437..122W}
{Wardle}, J.~F.~C., {Cawthorne}, T.~V., {Roberts}, D.~H., \& {Brown}, L.~F.
  1994, \apj, 437, 122

\bibitem[{{Zamaninasab} {et~al.}(2014){Zamaninasab}, {Clausen-Brown},
  {Savolainen}, \& {Tchekhovskoy}}]{2014Natur.510..126Z}
{Zamaninasab}, M., {Clausen-Brown}, E., {Savolainen}, T., \& {Tchekhovskoy}, A.
  2014, \nat, 510, 126

\bibitem[{{Zamaninasab} {et~al.}(2013){Zamaninasab}, {Savolainen},
  {Clausen-Brown}, {Hovatta}, {Lister}, {Krichbaum}, {Kovalev}, \&
  {Pushkarev}}]{2013MNRAS.436.3341Z}
{Zamaninasab}, M., {Savolainen}, T., {Clausen-Brown}, E., {et~al.} 2013,
  \mnras, 436, 3341

\bibitem[{{Zensus} {et~al.}(1995){Zensus}, {Cohen}, \&
  {Unwin}}]{1995ApJ...443...35Z}
{Zensus}, J.~A., {Cohen}, M.~H., \& {Unwin}, S.~C. 1995, \apj, 443, 35

\bibitem[{{Zhao} {et~al.}(2011){Zhao}, {Hong}, {An}, {Jiang}, {Zhao},
  {Gurvits}, \& {Yang}}]{2011A&A...529A.113Z}
{Zhao}, W., {Hong}, X.~Y., {An}, T., {et~al.} 2011, \aap, 529, A113

\end{thebibliography}
